\documentclass[aps,prd,nofootinbib,amssymb,eqsecnum,notitlepage]{revtex4-1}
\usepackage{graphicx}
\usepackage{bm}
\usepackage{amsmath,amsthm,amssymb}
\usepackage{color}

\usepackage{amsfonts}
\usepackage{hyperref}
\usepackage{here}

\begin{document}
\newcommand{\newc}{\newcommand}

\newc{\be}{\begin{equation}}
\newc{\ee}{\end{equation}}
\newc{\ba}{\begin{eqnarray}}
\newc{\ea}{\end{eqnarray}}
\newc{\bea}{\begin{eqnarray*}}
\newc{\eea}{\end{eqnarray*}}
\newc{\Mpl}{M_{\rm pl}}
\newc{\at}{\bar{a}_0}
\newcommand{\mm}[1]{\textcolor{red}{#1}}
\newcommand{\rkc}[1]{\textcolor{magenta}{[RK:~#1]}}
\newcommand{\mmc}[1]{\textcolor{blue}{[MM:~#1]}}
\newcommand{\stc}[1]{\textcolor{green}{[ST:~#1]}}
\allowdisplaybreaks[1]

\title{Relativistic stars in vector-tensor theories}

\author{Ryotaro Kase$^{1}$,
Masato Minamitsuji$^{2}$, and 
Shinji Tsujikawa$^{1}$}

\affiliation{$^1$Department of Physics, Faculty of Science, 
Tokyo University of Science, 1-3, Kagurazaka,
Shinjuku-ku, Tokyo 162-8601, Japan\\
$^2$Centro de Astrof\'{\i}sica e Gravita\c c\~ao  - CENTRA,
Departamento de F\'{\i}sica, Instituto Superior T\'ecnico - IST,
Universidade de Lisboa - UL, Av. Rovisco Pais 1, 1049-001 Lisboa, Portugal}

\date{\today}

\begin{abstract}
We study relativistic star solutions in second-order generalized 
Proca theories characterized by a $U(1)$-breaking vector field
with derivative couplings. 
In the models with cubic and quartic derivative coupling,
the mass and radius of stars 
become larger than those in general relativity 
for negative derivative coupling constants.
This phenomenon is mostly attributed to the increase of 
star radius induced by a slower decrease of the matter pressure compared to general relativity. 
There is a tendency that the relativistic star with a smaller mass 
is not gravitationally bound for a low central density and hence dynamically unstable,
but that with a larger mass is gravitationally bound.
On the other hand, we show that the intrinsic vector-mode couplings give rise to 
general relativistic solutions with a trivial field profile, 
so the mass and radius are not modified from those in general relativity.
\end{abstract}

\maketitle

\section{Introduction}
\label{sec1}

{}The increasing evidence of dark sectors in the Universe \cite{SNIa,CMB} 
implies that there may be some extra propagating degrees of freedom (DOFs) 
beyond the realm of General Relativity (GR). 
The new DOFs arising in modified gravitational 
theories can be potentially harmful as 
they generally mediate fifth forces with ordinary matter. 
In the local Universe with a weak gravitational field, however, there 
are several screening mechanisms of fifth forces known in the literature-- 
such as Vainshtein \cite{Vain} and chameleon \cite{chame} mechanisms. 
This screening property does not necessarily persist in 
the regime of strong gravity, reflecting the fact that 
the behavior of new DOFs can be modified by 
large nonlinearities in the field equations of motion.
The direct detections of gravitational waves by 
Advanced LIGO and Virgo \cite{LIGO,LIGO2} have 
opened up a new window for testing GR
in strong gravity regimes.

Besides black holes (BHs), relativistic stars are also 
important compact objects which allow one to
search possible deviation from GR in strong 
gravity regimes \cite{Vitor,Yagi}.
Especially, neutron stars (NSs)
are the representative relativistic stars.
Inside a NS, the gravitational force balances the degeneracy 
pressure of fermions \cite{Shapiro}. The properties of NSs, 
including the mass and radius, depend on the equation 
of state (EOS) of strong interacting matter, i.e., the relation between the matter pressure and 
density \cite{rev1,rev2}. 
The microscopic determination of the EOS of NSs 
from underlying nuclear interactions in an extremely high-density regime remains a challenging
theoretical problem. 

In modified gravitational theories, 
the existence of extra 
propagating DOFs can also influence the properties of relativistic 
stars. In scalar-tensor theories where a scalar field $\phi$ has 
a direct coupling with the Ricci scalar $R$, 
the Einstein-frame metric $g_{\mu\nu}$ felt 
by the matter sector is different from 
the Jordan-frame metric ${\tilde g}_{\mu\nu}$. 
The relation of these two metrics can be parametrized by the 
form ${\tilde g}_{\mu\nu}= A^2(\phi) g_{\mu\nu}$, 
where $A(\phi)$ is a function of $\phi$ \cite{Clifton:2011jh}. 
Inside a star, the conformal coupling to matter can trigger 
a tachyonic instability of the scalar field,
and spontaneously scalarizes the relativistic star.
Damour and Esposito-Far\`ese~\cite{Damour}
showed that such a scalarization, which occurs 
for the coupling e.g., $A(\phi) = \exp(\beta \phi^2/2)$ 
(where $\beta$ is a constant), significantly modifies 
the properties of relativistic stars with respect to GR.
Such a nontrivial excitation of the scalar field 
is a consequence of the absence of a no-hair theorem 
for stars. The scalarization can occur only for 
$\beta \lesssim -4.35$~\cite{Harada:1998ge,Novak:1998rk,Silva:2014fca},
whereas binary-pulsar observations~\cite{Freire:2012mg}
have set stringent bounds on $\beta$, as $\beta \gtrsim -4.5$.
For NSs, the existence of EOS-independent relations \cite{Doneva2017} 
will be important to resolve the degeneracies between the effects 
associated modified gravity and uncertainties in EOSs,
and test modified gravitational theories with future observations of NSs.

{}In shift-symmetric Horndeski (and beyond Horndeski) 
theories \cite{Galileon,Horndeski} with a minimally coupled 
matter component, the no-hair theorem for relativistic stars 
was argued in Ref.~\cite{Lehebel:2017fag}.
The theorem holds under the assumptions same 
as those used for proving the no-hair theorem of BHs in 
shift-symmetric Horndeski theories \cite{Hui}, 
with the regularity of metric functions 
and the scalar field at the center of stars.
Thus, as in the case of hairy 
BH solutions \cite{Soti1,Soti2,Babi14,Koba14,Lefteris,Babi16,Rinaldi:2012vy,Anabalon:2013oea,Minamitsuji:2013ura}, nontrivial NS configurations 
have been studied by violating at least one of those assumptions. 
For example, there exist relativistic star solutions for a linearly time-dependent 
scalar field $\phi=qt+\psi (r)$ \cite{Cisterna:2015yla, Cisterna:2016vdx,Maselli:2016gxk,Babichev:2016jom}.
Relativistic stars for other modified gravitational theories
have been extensively studied in Refs.~\cite{SVT,Quadratic,Massive}.
In this paper, we will study relativistic star solutions
in generalized Proca theories described by 
a $U(1)$-breaking vector field with derivative couplings.
We show that the star configuration  
with nontrivial influence of the extra 
DOFs can be constructed more easily 
compared to scalar-tensor theories. 

{}The action of generalized Proca theories with second-order 
equations of motion was first constructed in 
Refs.~\cite{Heisenberg,Tasinato} from the demand of keeping three propagating DOFs besides two tensor polarizations. The theories were further 
extended \cite{Allys} to include intrinsic vector-mode couplings with the double 
dual Riemann tensor $L^{\mu \nu \alpha \beta}$ \cite{Jimenez2016}, such that the $U(1)$-invariant interactions derived by Horndeski \cite{Horndeski76} can be accommodated as a specific case.
It is also possible to go beyond the second-order domain 
by keeping the five propagating 
DOFs \cite{HKT16,Kimura}.
In such (beyond) generalized Proca theories,
the derivative interactions can drive the late-time cosmic acceleration \cite{DeFelicecosmo} 
with some distinct observational signatures \cite{obsig1,obsig2}, while satisfying 
local gravity constraints in Solar System \cite{DeFeliceVain,Nakamura:2017lsf}. 

In the Einstein-Maxwell theory with a massless vector field,
the unique static and spherically symmetric BH solution 
corresponds to the Reissner-Nordstr\"{o}m (RN) metric with mass and electric charge. 
In the Einstein-Proca theory with a massive vector field
described by the Lagrangian  $-m^2 A^\mu A_\mu/2$,
Bekenstein showed that only the static and spherically symmetric 
BH solution is given by the Schwarzschild 
metric without the vector hair \cite{Bekenstein:1972ny}.
This no-hair theorem cannot be applied to vector-tensor theories 
with derivative self-interactions and nonminimal couplings to the spacetime curvature.
Indeed, it is known that there are a bunch of hairy 
BH solutions  in generalized Proca theories 
\cite{HorndeskiBH,Chagoya,Fan,Minami,Cisterna,Chagoya2,Babichev17,HKMT,HKMT2,Chagoya:2017ojn,Filippini:2017kov,Fan2017}.
In theories with a nonminimal coupling to the Einstein-tensor, $\beta_4 G^{\mu \nu} A_{\mu}A_{\nu}$,
Chagoya {\it et al.} \cite{Chagoya} derived an exact static and spherically symmetric BH solution for the 
specific coupling $\beta_4=1/4$.
This exact BH solution was further extended 
to asymptotically non-flat solutions \cite{Minami,Babichev17}, 
non-exact solutions for $\beta_4 \neq 1/4$ \cite{Babichev17,Chagoya2}, 
and slowly-rotating solutions \cite{Minami}. 
There are also exact BH solutions 
in a subclass of generalized Proca theories
with new internal symmetries \cite{Chagoya:2017ojn,Filippini:2017kov}.

In Refs.~\cite{HKMT,HKMT2}, analytic and numerical BH 
solutions have been systematically constructed 
for a wide class of generalized Proca theories.
The power-law coupling models,
which include the case of vector Galileons, 
can give rise to a variety of hairy BH solutions.
The cubic and quartic couplings 
provide BH solutions with a primary Proca hair, 
whereas the sixth-order and intrinsic vector-mode couplings
lead to BH solutions with a secondary Proca hair.
On the other hand, there are no regular BHs
for quintic power-law couplings
due to the divergence of the longitudinal mode 
at a finite radius.

While both BHs and stars are compact objects with strong gravitational 
forces, the internal structures of them are different. 
For static and spherically symmetric BHs the metric and curvature 
generally exhibit the divergence at the center of spherical symmetry, 
but this is not the case for stars.  
Moreover, the configuration of stars is affected by 
different choices of the EOS  through the change of the 
matter pressure. In this paper, we will study how 
the presence of derivative couplings in generalized Proca theories 
affects the mass and radius of relativistic stars. 
In Ref.~\cite{Chagoya2}, the authors studied NS solutions 
in a subclass of generalized Proca theories
with the Lagrangian $\beta_4 G^{\mu \nu} A_{\mu}A_{\nu}$.
We extend the analysis to more general cubic and quartic 
power-law derivative couplings and elucidate general properties of their effects on the mass and radius of relativistic stars. 

For our purpose of investigating the effects of cubic and quartic derivative 
couplings on the mass and radius of relativistic stars in comparison with GR,
we will restrict our numerical analysis to the case of the simplest 
polytropic EOS with two constant parameters \cite{Baum}.
We derive analytic solutions deep inside the star by imposing 
regular boundary conditions at the origin. 
The validity of analytic solutions will be confirmed by  
numerical integrations across the surface of star for the polytropic EOS.
We will also study the effects of sixth-order and intrinsic vector-mode 
couplings on the configuration of relativistic stars.
However, we will not consider quintic derivative couplings 
because of the absence of regular BHs \cite{HKMT,HKMT2} 
as well as pathological behavior
in the regime of weak gravity \cite{Nakamura:2017lsf}. 
The essential qualitative features of relativistic stars in generalized 
Proca theories are not sensitive to the choice of EOSs.

We organize our paper as follows.
In Sec.~\ref{sec2}, we derive a set of equations 
in generalized Proca theories 
with matter on the static and spherically symmetric background, and briefly review relativistic stars in GR and the polytrope EOS. In Secs.~\ref{sec4} and \ref{sec5}, 
we study how the mass and radius 
are modified by the presence of cubic and quartic power-law couplings, 
respectively.
In Sec.~\ref{sec6}, we show that sixth-order and intrinsic vector-mode couplings lead to the relativistic star solutions identical to those in GR with a trivial vector field. 
We conclude in Sec.~\ref{sec7}.

We work in the CGS units, where the speed of light, the reduced Planck constant, 
the gravitational constant, 
and the neutron mass are given by 
$c=2.9989 \times 10^{10}$ cm$\cdot$s$^{-1}$,
$\hbar=1.0546 \times 10^{-27}$~erg$\cdot$s,
$G=6.6741 \times 10^{-8}$~g$^{-1}$$\cdot$${\rm cm}^3$$\cdot$s$^{-2}$,
and $m_{\rm n}=1.6749 \times 10^{-24}$ g,
respectively.

\section{Generalized Proca theories and relativistic stars}
\label{sec2}

\subsection{Equations of motion on the static and spherically symmetric background}
\label{sec2a}

The action of generalized Proca theories with a vector field $A_{\mu}$ is given by  \cite{Heisenberg,Jimenez2016}
\be
S=\int d^{4}x \sqrt{-g} 
\left[
F
+\sum_{i=2}^{6} \mathcal{L}_{i}
+\mathcal{L}_m 
\right]\,,
\label{action}
\ee
where $g$ is a determinant of the metric tensor
$g_{\mu \nu}$, ${\cal L}_m$ is a matter Lagrangian, and
\ba
\mathcal{L}_{2}&=& G_{2}(X, F, Y)\,,
\label{L2}
\\
\mathcal{L}_{3}&=& G_{3}(X) \nabla_{\mu} A^{\mu}\,,
\label{L3}
\\
\mathcal{L}_{4}&=& G_{4}(X) R 
+ G_{4,X}(X)\left[ (\nabla_{\mu} A^{\mu})^{2} 
-  \nabla_{\mu} A_{\nu} \nabla^{\nu} A^{\mu}\right] \,,\\
\mathcal{L}_{5}&=& G_{5}(X) G_{\mu\nu} \nabla^{\mu} A^{\nu} 
- \frac{1}{6} G_{5,X} (X) \left[ (\nabla_{\mu} A^{\mu})^{3} 
- 3 \nabla_{\mu} A^{\mu} \nabla_{\rho} A_{\sigma} \nabla^{\sigma} A^{\rho} 
+ 2 \nabla_{\rho} A_{\sigma} \nabla^{\nu} A^{\rho} \nabla^{\sigma} A_{\nu} \right]
\notag
\\
&&-g_{5}(X) \tilde{F}^{\alpha\mu}\tilde{F}^{\beta}_{~\mu} \nabla_{\alpha} A_{\beta}\,,
\\
\mathcal{L}_{6}&=& G_{6}(X) L^{\mu\nu\alpha\beta} \nabla_{\mu} A_{\nu} \nabla_{\alpha} A_{\beta}
+\frac{1}{2} G_{6,X}(X) \tilde{F}^{\alpha\beta} \tilde{F}^{\mu\nu} \nabla_{\alpha} 
A_{\mu} \nabla_{\beta} A_{\nu}\,,
\label{L6}
\ea
with 
\ba
&&F_{\mu \nu}=\nabla_{\mu}A_{\nu}-\nabla_{\nu}A_{\mu}\,,\qquad 
F=-\frac{1}{4}F_{\mu \nu}F^{\mu \nu}\,,\qquad 
X=-\frac{1}{2}A_{\mu}A^{\mu}\,,\qquad 
Y=A^{\mu}A^{\nu}{F_{\mu}}^{\alpha}F_{\nu \alpha}\,.
\ea
Here, $\nabla_\mu$, $R$, and $G_{\mu\nu}$ 
represent the covariant derivative, 
the Ricci scalar, and the Einstein tensor
associated with the four-dimensional 
metric $g_{\mu\nu}$, respectively.
While the function $G_2$ is generally dependent on 
$X, F, Y$,  the functions $G_{3,4,5,6}$ and $g_5$ 
depend on $X$ alone
with the notation of partial derivatives
$G_{i,X} \equiv \partial G_{i}/\partial X$.
The dual strength tensor $\tilde{F}^{\mu\nu}$ and 
the double dual Riemann tensor 
$L^{\mu \nu \alpha \beta}$ are defined, respectively, by 
\be
\tilde{F}^{\mu\nu}=\frac{1}{2} \mathcal{E}^{\mu\nu\alpha\beta} F_{\alpha\beta}\,,\qquad
L^{\mu\nu\alpha\beta}=\frac{1}{4} \mathcal{E}^{\mu\nu\rho\sigma} \mathcal{E}^{\alpha\beta\gamma\delta} R_{\rho\sigma\gamma\delta}\,,
\ee
where $\mathcal{E}^{\mu\nu\alpha\beta}$ is the Levi-Civita tensor satisfying the normalization  
$\mathcal{E}^{\mu\nu\alpha\beta}\mathcal{E}_{\mu\nu\alpha\beta}=-4!$, and 
$R_{\rho\sigma\gamma\delta}$ is the Riemann tensor. 
The Lagrangians containing the functions $g_5(X)$ and  
$G_6(X)$ correspond to intrinsic vector-modes.

We consider a static and spherically symmetric background 
characterized by the line element 
\be
ds^{2} =-f(r)
c^2 dt^{2} +h^{-1}(r)dr^{2} + 
r^{2} \left( d\theta^{2}+\sin^{2}\theta\, 
d \varphi^{2} \right)\,,
\label{metric}
\ee
where $f$ and $h$ are functions of the distance $r$ from 
the center of symmetry.
On this background, the vector field can be expressed 
in the form 
\be
A_{\mu}=\left( c A_0(r), A_1(r), 0, 0 \right)\,,
\label{vector_ansatz}
\ee
where $A_1(r)$ is the $r$-derivative of a longitudinal 
scalar $\chi$, such that $A_1(r)=d\chi/dr \equiv \chi'(r)$. 
The transverse mode $A_i^{(T)}$ in the spatial components 
$A_i$ needs to vanish due to the regularity 
at the origin \cite{DeFeliceVain}. 
On the static and spherically symmetric background (\ref{metric}) with the 
vector components (\ref{vector_ansatz}) there is the relation 
$Y=4FX$, so the additional dependence of $Y$ in Eq.~(\ref{L2}) can be removed \cite{HKMT2}.

We assume that the matter sector is described by 
a perfect fluid minimally coupled to gravity. 
Defining the matter energy-momentum tensor
\be
T^{\mu\nu} \equiv \frac{2}{\sqrt{-g}} \frac{\delta (\sqrt{-g} {\cal L}_m)}{\delta g_{\mu\nu}}\,,
\ee
the mixed tensor $T^{\mu}_{\nu}$ is expressed 
in the form 
\be
T^{\mu}_{\nu}={\rm diag} \left( -\rho c^2, 
P, P, P \right)\,,
\ee
where $\rho$ is the total mass density 
and $P$ is the pressure.

Varying the action (\ref{action}) with respect to 
$f, h, A_0, A_1$, respectively, we obtain 
\ba
& &
\left( c_{1} + \frac{c_{2}}{r} + \frac{c_{3}}{r^{2}} \right) h' 
+ c_{4} + \frac{c_{5}}{r} + \frac{c_{6}}{r^{2}}
=\frac{\rho}{c^2}\,, 
\label{be1} \\
& &-\frac{h}{f} \left( c_{1} + \frac{c_{2}}{r} + \frac{c_{3}}{r^{2}} \right) f' 
+ c_{7} + \frac{c_{8}}{r} + \frac{c_{9}}{r^{2}}
=\frac{P}{c^4}
\,,
\label{be2}
\ea
where $c_{1,2,\cdots,9}$ are given in Appendix \ref{app}, and 
\ba
\hspace{-0.4cm}
& & 
rf \left[ 2fh(rA_0''+2A_0')+r(fh'-f'h)A_0' \right] (1+G_{2,F})
+r^2hA_0'^2 \left[ 2fhA_0''-(f'h-fh') A_0' \right] G_{2,FF}
-2 r^2f^2A_0 G_{2,X}
\notag\\
\hspace{-0.4cm}
& & 
-2 r^2fA_0' \left( fh^2A_1A_1' -hA_0A_0'+f'hX_0-fh'X_1\right) G_{2,XF}
-rfA_0 \left[ 2 rfhA_1'+(rf'h+rfh'+4fh)A_1 \right] G_{3,X}
\notag\\
\hspace{-0.4cm}
& & 
+4 f^2A_0 (rh'+h-1) G_{4,X}
-8 fA_0 \left[ rfh^2 A_1A_1'-(rf'h+rfh'+fh) X_1\right] G_{4,XX}
\notag\\
\hspace{-0.4cm}
& & 
-fA_0 \left[ f(3h-1)h'A_1+h(h-1) (f'A_1+2fA_1')   \right] G_{5,X}
-2 fhA_0X_1\left[2 fhA_1'+(f'h+fh')A_1 \right] G_{5,XX}
\notag\\
\hspace{-0.4cm}
& & 
-2 f \left[ f (3 h-1) h'A_0'+h (h-1) (2fA_0''-f'A_0') \right] G_6
-4 fh A_0'X_1 \left( h A_0 A_0'-2 fh^2 A_1 A_1'
-2 f'hX_0+2 fh'X_1 \right) G_{6,XX}
\notag\\
\hspace{-0.4cm}
& & 
-2 f \left[ 4 fh^2 X_1 A_0''-2 h (hX- X_0) f'A_0'
+2f (6h-1)h' X_1A_0'+h(h-1)A_0A_0'^2
-2 fh^2(3h-1) A_0'A_1A_1'\right] G_{6,X}
\notag\\
\hspace{-0.4cm}
& & 
-4 fh \left[ 2 rfh A_1 A_0''- \{(rf' h-3 rfh'-2fh) A_1-2 rfhA_1'\} A_0'  \right] g_5
\notag\\
\hspace{-0.4cm}
& & 
-4 rfh A_0' \left[ hA_0A_0'A_1+4 fhX_1A_1'-2 A_1(f'hX_0-fh'X_1)\right] g_{5,X}
=0\,,
\label{be4}\\
\hspace{-0.4cm}
& & 
A_1 \left[ r^2fG_{2,X}-2 (rf'h+fh-f) G_{4,X}
+4h(rA_0 A_0'-rf' X-fX_1) G_{4,XX}
-hA_0'^2(3h-1) G_{6,X}-2h^2X_1 A_0'^2 G_{{6,{  XX}}}\right]
\nonumber \\
\hspace{-0.4cm}
& &=r [r(f' X-A_0 A_0')+4 fX_1] G_{3,X}
+2 f'hX_1G_{5,X}+(A_0 A_0'-f' X)\left[ (1-h)G_{5,X}-2 hX_1G_{5,XX}\right]
\nonumber \\
\hspace{-0.4cm}
& &
\hspace{.35cm}
-2rh A_0'^2( g_{5} +2 X_1 g_{5,X})\,.
\label{be5}
\ea
The quantity $X$ is given by $X=X_0+X_1$, where
\be
X_0  \equiv \frac{A_0^2}{2f}\,,\qquad 
X_1 \equiv -\frac{hA_1^2}{2}\,.
\ee
{}From the matter continuity equation, it follows that 
\be
P'+\frac{f'}{2f} \left( \rho c^2+P \right)=0\,.
\label{mattereq}
\ee
For a given EOS 
\be
\label{eos2}
P=P (\rho)\,,
\ee
Eqs.~(\ref{be1})-(\ref{be5}) with Eq.~(\ref{mattereq})
form a closed set of equations
to determine $f,h,A_0,A_1, \rho$, and $P$ 
as functions of $r$.

\subsection{Relativistic stars in GR}
\label{sec2b}

Here, we briefly review relativistic stars in GR without the vector field $A_{\mu}$.
This corresponds to the functions
\be
G_4=\frac{1}{16\pi G}\,,\qquad 
G_2=G_3=G_5=G_6=0\,,\qquad g_5=0\,.
\ee
In this case, Eqs.~(\ref{be1}) and (\ref{be2}) reduce, 
respectively, to 
\ba
& &
\frac{h'}{r}+\frac{h-1}{r^2}=-\frac{8\pi G \rho}{c^2}\,,
\label{GR1}\\
& &
\frac{h}{f}\frac{f'}{r}+
\frac{h-1}{r^2}=\frac{8\pi G P}{c^4}\,.
\label{GR2}
\ea
Introducing the mass function $M(r)$, as 
\be
h(r)=1-\frac{2GM(r)}{c^2r}\,,
\label{hr}
\ee
we can express Eq.~(\ref{GR1}) in the simple form
\be
M'(r)=4\pi \rho r^2\,.
\label{GR3}
\ee
On using Eqs.~(\ref{GR2}) and (\ref{hr}), the continuity 
equation (\ref{mattereq}) reduces to the 
Tolman-Oppenheimer-Volkoff (TOV) equation 
\be
P'(r)=-\frac{G(\rho+P/c^2)(M+4\pi r^3 P/c^2)}
{r^2[1-2GM/(c^2 r)]}\,.
\label{GR4}
\ee

Around the center of star, we expand $f,h,\rho$, and $P$ 
in the following forms
\be
f(r)=1+\sum_{i=2}^{\infty} f_i r^i\,,\qquad
h(r)=1+\sum_{i=2}^{\infty} h_i r^i\,,\qquad
\rho(r)=\rho_c+\sum_{i=2}^{\infty} \rho_i r^i\,,\qquad
P(r)=p_c+\sum_{i=2}^{\infty} p_i r^i\,,
\label{fh}
\ee
where $f_i,h_i,\rho_i,p_i$ are constants.
Then, the regularity conditions 
$f'(0)=h'(0)=\rho'(0)=P'(0)=0$ are satisfied with 
$\rho(r)$ and $P(r)$ converging to constant values 
$\rho_c$ and $p_c$, respectively, as $r \to 0$. 
By solving Eqs.~\eqref{GR1}, \eqref{GR2}, and \eqref{mattereq} iteratively,
the boundary conditions around $r=0$ can be found as 
\ba
&&f(r)=1+\frac{4\pi G(c^2\rho_c+3p_c)}{3c^4} r^2+{\cal O}(r^4)\,, \label{gr1}\\
&&h(r)=1-\frac{8\pi G\rho_c}{3c^2} r^2+{\cal O}(r^4)\,,\label{gr2} \\
&&P(r)=p_c
-\frac{2\pi G(c^2\rho_c+3p_c)(c^2\rho_c+p_c) }{3c^4}r^2
+{\cal O}(r^4)\,.\label{gr3}
\ea
The numerical integration is performed until reaching the surface of star $r=R_\ast$,
where $P(R_\ast)=0$.
By requiring the continuity of metric functions 
and their first-order derivatives across the 
surface $r=R_\ast$, the internal solution is smoothly 
joined to the exterior Schwarzschild solution
given by the metric \eqref{metric} with
\be
f=h= 1-\frac{2GM_{\ast}}{c^2 r},
\label{Sch}
\ee
where the Arnowitt-Deser-Misner (ADM) mass 
is given by $M_{\ast} \equiv M(R_{\ast})$.
Provided that the EOS (\ref{eos2}) inside the star is known,
it is practically more convenient to integrate 
Eqs.~(\ref{GR2}), (\ref{GR3}), and (\ref{GR4}) to 
determine $\rho(r), P(r)$ and $M(r)$. 
In Secs.~\ref{sec4}-\ref{sec6}, the mass and radius of relativistic stars in generalized Proca theories will be 
compared to those in GR.

\subsection{The polytrope equation of state}
\label{sec2c}

As we will see later, the qualitative results of relativistic stars 
in generalized Proca theories do not depend on 
the choice of EOSs. Thus, in this paper, we focus on one 
of the simplest EOSs known as the polytropic EOS 
which is given by 
\be
P={\cal K}\rho_0^{\Gamma}\,,
\label{poly}
\ee
where $\rho_0$ is the rest-mass density, and 
${\cal K}$, $\Gamma$ are constants.
In general, the total energy density $\rho c^2$ is expressed 
in the form $\rho c^2=\rho_0 c^2 (1+\epsilon)$, 
where $\epsilon$ 
is the dimensionless internal energy density per unit mass.
For baryons with number density $n_b$
and the mean rest mass $m_b$,
the rest-mass density is given by $\rho_0=n_bm_b$.
On using the first law of thermodynamics for the 
adiabatic process, the baryon pressure is expressed as
$P=n_b^2 m_bc^2\,\partial \epsilon/\partial n_b$ \cite{Baum}. 
For the polytropic EOS (\ref{poly}), i.e., 
$P={\cal K}(n_bm_b)^{\Gamma}$, we obtain the 
integrated solution 
$\epsilon={\cal K}\rho_0^{\Gamma-1}/[c^2 (\Gamma-1)]$, 
so the total mass density yields 
$\rho=\rho_0+{\cal K}\rho_0^{\Gamma}/[c^2 (\Gamma-1)]$. 
We define the dimensionless rest-mass density $\chi$ and 
the rescaled polytropic gas constant $K$, as 
\be
\chi \equiv \frac{\rho_0}{\tilde{\rho}_0}=\frac{n_b}{n_0}
\,,\qquad
K \equiv \frac{{\cal K}}{\tilde{\rho}_0^{1-\Gamma}c^2}\,,
\ee
with 
\be
\tilde{\rho}_0=n_0m_b\,,
\ee
where $n_0=0.1\,{\rm (fm)^{-3}}$ is the typical nuclear 
number density of relativistic stars. 
As a result, the polytropic EOS can be expressed 
in the form \cite{Damour}
\be
\rho=\tilde{\rho}_0 \left( \chi+\frac{K}{\Gamma-1}
\chi^{\Gamma} \right)\,,\qquad 
P=K\tilde{\rho}_0c^2 \chi^{\Gamma}\,,
\label{eos}
\ee
with 
\be
w \equiv \frac{P}{\rho c^2}
=\frac{K \chi^{\Gamma-1}}
{1+K\chi^{\Gamma-1}/(\Gamma-1)}\,.
\label{wpoly}
\ee
In the nonrelativistic regime characterized by 
$K \chi^{\Gamma-1} \ll 1$, we have 
$w \simeq K \chi^{\Gamma-1}$, so $w$ grows 
with the increase of mass density $\rho$. 
In the relativistic regime, $w$ approaches 
a constant value $\Gamma-1$ for increasing $\rho$.

For the numerical propose, it is convenient to introduce 
the dimensionless quantities: 
\be
x=\frac{r}{r_0}\,,\qquad 
y=\frac{\rho}{\tilde{\rho}_0}\,,\qquad 
w_0=\frac{P}{\tilde{\rho}_0 c^2}\,,\qquad 
m(r)=\frac{3M(r)}{4\pi \tilde{\rho}_0 r_0^3}\,,
\label{wdef}
\ee
where 
\be
r_0=\sqrt{\frac{c^2}{G\tilde{\rho}_0}}\,.
\label{r0def}
\ee
In the following, we identify $m_b$ with the neutron mass 
$m_{\rm n}=1.6749 \times 10^{-24}$~g. 
Then, the distance (\ref{r0def}) corresponds to 
$r_0=89.696$~km with 
$\tilde{\rho}_0=1.6749 \times 10^{14}$ g~cm$^{-3}$.
The polytropic EOS (\ref{eos}) can be 
expressed in the form 
\be
w_0(x)=K \chi(x)^{\Gamma}\,,\qquad 
y(x)=\left( \frac{w_0(x)}{K} \right)^{1/\Gamma}
+\frac{w_0(x)}{\Gamma-1}\,. 
\ee
Specifying the value of $w_0(0)$, 
the associated dimensionless density 
$y_c=\rho_c/\tilde{\rho}_0$ 
is also fixed at the center of star. 
The star radius $R_{\ast}$ is defined by 
\be
w_0(x_*)=0\,,
\ee
where $x_{\ast}=R_{\ast}/r_0$.
By choosing different boundary conditions of $w_0$ at 
$x=0$, we obtain the configuration of relativistic stars
with different mass $M_{\ast}$ and radius $R_{\ast}$. 
In terms of the solar mass 
$M_{\odot}=1.9884 \times 10^{33}$ g, we can 
express the ADM mass $M_{\ast}$ in the form
\be
M_{\ast}=2.5462 \times 10^2\,m(x_{\ast}) M_{\odot}\,.
\ee

For the comparison with observational data of NSs, 
however, we would need phenomenologically parametrized 
EOSs specifying the stiffness of the star in several 
density intervals \cite{Read2008}.
In this paper, we will not perform 
the comparison with observational data of NSs,
but we focus on how vector-field derivative couplings modify 
the mass-radius relation of relativistic stars from GR
by considering the polytropic EOS (\ref{eos})
with two constant parameters $\Gamma$ and $K$.
As we will see below, the qualitative behavior of 
vector-field derivative couplings on the mass and radius
of relativistic stars, which can be analytically  
understood in some degree, is generally insensitive to the 
choice of EOSs. For numerics, we choose the index 
$\Gamma=2.34$ in Secs.~\ref{sec4} and \ref{sec5}.

\section{Cubic couplings}
\label{sec4}

Let us begin with the cubic derivative interaction $G_3(X)$. 
For concreteness, we study the power-law coupling given by  
\be
G_3=\beta_3 X^n\,,
\label{cuG3}
\ee
where $\beta_3$ is a constant 
and $n$ is a positive integer.
We also take into account the Einstein-Hilbert term 
$G_4=1/(16\pi G)$ in the action (\ref{action}) 
with $G_2=G_5=G_6=0$ and $g_5=0$. 
We consider the models with positive integer $n$, which 
includes the vector Galileon as a special case ($n=1$).
From Eq.~(\ref{be5}), the longitudinal component 
is related to $A_0,f,h$ and their derivatives as 
\be
A_1=\epsilon
\sqrt{\frac{rA_0 (f'A_0-2fA_0')}
{fh (rf'+4f)}}\,,
\label{A1be3}
\ee
where $\epsilon=\pm 1$.

\subsection{Analytic solutions around the center of star}

We first derive analytic solutions to the metrics, the vector 
field, and the pressure around $r=0$.
We take the positive branch of Eq.~(\ref{A1be3}) and 
differentiate it with respect to $r$. 
Then, $A_1$ and $A_1'$ are substituted into 
Eqs.~(\ref{be1})-(\ref{be4}) to eliminate 
the dependence of the longitudinal mode.

Around the center of star, we expand $f,h,\rho,P$ 
in the forms (\ref{fh}). The temporal vector 
component is also expanded as 
\be
A_0=a_0+\sum_{i=2}^{\infty} a_i r^i\,,
\label{fhA0}
\ee
where $a_0$ and $a_i$ are constants. 
These solutions satisfy the regular boundary 
conditions $f'(0)=h'(0)=\rho'(0)=P'(0)=0$ and $A_0'(0)=0$. 
Without loss of generality, we will assume that $a_0>0$. 
We also require the condition $P''(0)<0$ for 
the pressure \cite{Delgaty}. 
Expanding the continuity equation (\ref{mattereq}) 
around the origin, we obtain 
\be
p_2=-\frac{\rho_c c^2+p_c}{2}f_2\,.
\label{p2}
\ee
The condition $P''(0)<0$, which corresponds to $p_2<0$, 
is satisfied for 
\be
f_2>0\,. 
\label{matcon}
\ee
{}From Eq.~(\ref{A1be3}), the leading-order solution of 
the longitudinal mode around the center of star
is given by 
\be
A_1=\sqrt{\frac{a_0 (a_0 f_2-2a_2)}{2}}r\,, 
\label{A1ho}
\ee
which ensures the regularity of $A_1$ at $r=0$.
For the existence of this solution, we require that  
\be
a_0(a_0f_2-2a_2)>0\,. 
\label{G3con1}
\ee

Substituting Eq.~(\ref{fhA0}) into 
Eqs.~(\ref{be1})-(\ref{be4}) 
and solving them iteratively, we obtain the following solutions around the origin: 
\ba
&&f(r)=1+\frac{4\pi}{3} \left(1+3w_c+{\cal F} \right) 
\frac{r^2}{r_c^2}+{\cal O}(r^4)\,,\label{G3ini1} \\
&&h(r)=1-\frac{8\pi}{3} \left( 1+{\cal F} \right)
\frac{r^2}{r_c^2}+{\cal O}(r^4)\,,\label{G3ini2}\\
&&A_0(r)=\frac{\bar{a}_0}{\sqrt{8\pi G}} \left( 
1+\frac{4\pi}{3} \frac{{\cal F}}{\bar{a}_0^2}
\frac{r^2}{r_c^2}\right)+{\cal O}(r^4)\,,
\label{G3ini3} \\
&&P(r)=p_c
-\frac{2\pi}{3} \left(\rho_c c^2+p_c \right)
\left(1+3w_c+{\cal F} \right) 
\frac{r^2}{r_c^2}+{\cal O}(r^4),
\label{G3ini4}
\ea
where 
\be
{\cal F} \equiv 
\frac{3n^2\bar{a}_0^{2n+1}\bar{\beta}_3}
{2^{2n+3} \pi}
\left[ -\bar{\beta}_3 \at^{2n-1} \left(1-\frac{\at^2}{2}\right)\pm \sqrt{\bar{\beta}_3^2\at^{4n-2}
\left(1-\frac{\at^2}{2}\right)^2
+\frac{2^{2n+3}\pi}{3n^2}(1+3w_c)}\right]\,,
\label{defF}
\ee
with the dimensionless constants defined by 
\be
\bar{\beta}_3 \equiv 
\frac{\beta_3 r_c}{(\sqrt{8 \pi G})^{2n-1}}\,,
\qquad \at \equiv 
\sqrt{8 \pi G} a_0\,,\qquad 
r_c \equiv 
\sqrt{\frac{c^2}{G\rho_c}}\,,\qquad 
w_c \equiv \frac{p_c}{\rho_c c^2}\,.
\label{wcdef}
\ee
In the limit that $\beta_3 \to 0$, the iterative solutions 
(\ref{G3ini1}), (\ref{G3ini2}), and (\ref{G3ini4}) recover 
the general relativistic solutions 
(\ref{gr1}), (\ref{gr2}), and (\ref{gr3}), respectively. 
The density $\rho(r)$ is known for a given EOS.
On using Eq.~(\ref{G3ini1}) with 
Eq.~(\ref{defF}), the condition (\ref{matcon}) 
translates to 
\be
|\bar{\beta}_3| \bar{a}_0^{2n} 
< \frac{2^{n+1}}{n} 
\sqrt{\frac{\pi (1+3w_c)}{3}}\,.
\label{bebo1}
\ee
Under this bound, the condition (\ref{G3con1}) is automatically satisfied. 
The EOS $w_c$ is bounded from above with the 
maximum value of order 1. 
For the polytropic EOS (\ref{wpoly}), 
we have that $w_c <\Gamma-1$. 
Then, $|\bar{\beta}_3| \bar{a}_0^{2n} \lesssim 2^{n+1}/n$ 
from Eq.~(\ref{bebo1}). 
For $n={\cal O}(1)$, the product $|\bar{\beta}_3| \bar{a}_0^{2n}$ is constrained to be smaller than the order of 1.

For the branch of the positive sign in Eq.~(\ref{defF}) 
the upper bound (\ref{bebo1}) corresponds to the negative value of $\bar{\beta}_3$, whereas, for the negative sign, the upper limit of $\bar{\beta}_3$ is positive.
In the following, we will focus on the case of positive sign 
in Eq.~(\ref{defF}) without loss of generality. 
Then, for $\bar{\beta_3}<0$, the term 
${\cal F}$ in Eq.~(\ref{G3ini4}) is negative, 
so the negative coupling $\bar{\beta}_3$ effectively increases 
the pressure. In other words, the positive term 
$1+3w_c$ in Eq.~(\ref{G3ini4}) is partially compensated 
by the negative term ${\cal F}$. 
This means that, with increasing $r$, the pressure
$P(r)$ decreases more slowly relative to the case $\bar{\beta}_3=0$ 
at least around the center of body.
Then, we expect that the negative coupling $\bar{\beta}_3$
may lead to a larger radius of star than that 
for $\bar{\beta}_3=0$.

Indeed, the negative value of $\bar{\beta}_3$ close to the upper bound 
of Eq.~(\ref{bebo1}) gives rise to the pressure (\ref{G3ini4}) which is 
nearly constant around the center of star. 
Then, we may anticipate that the radius of star can be 
infinitely large. However, we will show that this is 
not the case. {}From Eqs.~(\ref{hr}) and (\ref{G3ini2}) 
the mass function around $r=0$ is given by 
\be
M(r)=\frac{4}{3}\pi \rho_c r^3
\left(1+{\cal F} \right)+{\cal O}(r^5)\,.
\label{Mass}
\ee
The negative coupling $\bar{\beta}_3$ leads to the 
decrease of $M(r)$ relative to the case of GR.
For the theoretical consistency, we require that 
$M(r)>0$ around the center of body. 
This amounts to the condition ${\cal F}>-1$, 
which translates to 
\be
|\bar{\beta}_3| \bar{a}_0^{2n} <
\frac{2^{n+1}}{n}\sqrt{\frac{2\pi}
{3(2+3w_c\bar{a}_0^2)}}\,,
\label{bebo3}
\ee
which is tighter than the bound (\ref{bebo1}).
Substituting ${\cal F}=-1$ into Eq.~(\ref{G3ini4}), the 
pressure corresponding to the maximum value 
of $|\bar{\beta}_3|$ in Eq.~(\ref{bebo3})
is given by 
\be
P_{\rm max}(r)=p_c \left[ 1-2\pi (1+w_c) 
\frac{r^2}{r_c^2} \right]\,,
\label{Pmax}
\ee
which decreases for increasing $r$. 
This expression is valid around $r=0$, but we extrapolate it to the surface of star to provide a crude criterion 
for the upper limit of the radius $R_{\ast}$. Then, we obtain the bound 
\be
R_{\ast} \lesssim \frac{r_c}{\sqrt{2\pi (1+w_c)}}\,,
\label{Rstar}
\ee
which means that $R_{\ast}$ is constrained to be 
smaller than the order of $r_c$. 
Since the r.h.s. of Eq.~(\ref{Rstar}) does not depend on 
the power $n$, the maximum radius is insensitive to 
the form of cubic couplings $G_3(X)$. 

To discuss the dynamical stability of relativistic stars, 
we define the proper mass 
\be
M_p \equiv \int_{R \le R_{\ast}} d^3x\, \rho 
\sqrt{{}^{(3)}g}
=\int_0^{R_{\ast}} \frac{4\pi \rho\,r^2}{\sqrt{h}}dr\,,
\label{Mp}
\ee
where ${}^{(3)}g$ is the determinant of 
three-dimensional spatial metric. 
The gravitational binding energy is defined by
the difference between $M_p$ and 
the ADM mass $M_\ast$, i.e.,
\be
\Delta \equiv \left( M_p-M_* \right)c^2\,.
\label{Delta}
\ee
The star with $\Delta>0$ is gravitationally bound
and the condition $\Delta>0$ can be regarded as 
a necessary condition for its dynamical stability,
whereas the star with $\Delta<0$ is not bound 
and hence dynamically unstable.
For $\bar{\beta}_3<0$ the $r$-derivative of the leading-order 
term on the r.h.s. of Eq.~(\ref{Mass}) 
is smaller than $4\pi \rho_c r^2$, whereas the term inside the 
integral of Eq.~(\ref{Mp}) is larger than $4\pi \rho\,r^2$. 
This implies that the condition $\Delta>0$ may hold 
for $\bar{\beta}_3<0$, but we need to caution that 
Eq.~(\ref{Mass}) is valid only around the central 
region of star.

\subsection{Numerical solutions}

The above analytic solutions have been derived under the expansion around $r=0$. 
In order to study the effect of the coupling $\beta_3$ on 
the mass $M_{\ast}$ and the radius $R_\ast$
of relativistic stars more precisely, we numerically solve 
Eqs.~(\ref{be1})-(\ref{be5}) with Eq.~(\ref{mattereq}) 
for the polytropic EOS (\ref{eos}) by using the boundary conditions (\ref{G3ini1})-(\ref{G3ini4}) around the origin. 
For numerical computations, we will focus on the 
case of vector Galileons, i.e., $n=1$ in Eq.~(\ref{cuG3}).
The numerical integration is performed 
until reaching the surface $r=R_{\ast}$ characterized by the 
condition $w_0(R_{\ast})=0$, where $w_0$ is defined 
in Eq.~(\ref{wdef}).
By requiring the continuity of metric functions, 
the vector field, and their first-order derivatives 
across the surface $r=R_{\ast}$ and using their values 
at $r=R_{\ast}$ as boundary conditions, 
the exterior solution can be obtained by integrating 
Eqs.~(\ref{be1})-(\ref{be5}) in the vacuum region 
$r>R_{\ast}$, where $\rho=P=0$.
The consistent exterior solutions of star
approach the iterative solutions in the large $r$ limit
characterized by three parameters including the ADM mass $M_\ast$ 
(see Eqs.~(5.10)-(5.13) of Ref.~\cite{HKMT2}).

\begin{figure}[h]
\begin{center}
\includegraphics[height=3.2in,width=3.4in]{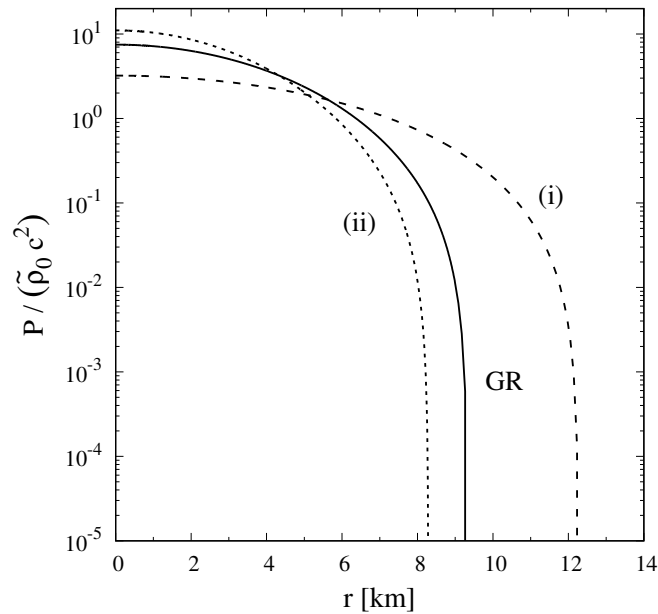}
\end{center}
\caption{\label{fig1}
Variation of the pressure in cubic Galileons 
($G_3=\beta_3X$) for the 
polytropic EOS (\ref{eos}) with $K=0.0130$ and $\Gamma=2.34$. 
The two cases (i) and (ii) correspond to 
(i) $\tilde{\beta}_3=-1$, 
$\bar{a}_0=2.2$, $\chi(r)=10.471$ at $r/r_0=10^{-3}$ and 
(ii) $\tilde{\beta}_3=1$, 
$\bar{a}_0=2.0$, $\chi(r)=17.783$ at $r/r_0=10^{-3}$, 
respectively. We also show the case of GR 
with $\tilde{\beta}_3=0$, 
$\bar{a}_0=0$, $\chi(r)=15.136$ at $r/r_0=10^{-3}$. 
The boundary conditions of $f,h,A_0,P$ are chosen 
to be consistent with Eq.~(\ref{G3ini1})-(\ref{G3ini4}). 
}
\end{figure}

In Fig.~\ref{fig1}, we plot the normalized pressure 
$P/(\tilde{\rho}_0c^2)$ 
versus the distance $r$ from the center of star 
with $K=0.013$ and $\Gamma=2.34$ for three different 
values of $\tilde{\beta}_3 \equiv \beta_3 r_0/\sqrt{8\pi G}
=\bar{\beta}_3 r_0/r_c$. 
In GR, the pressure varies according 
to Eq.~(\ref{gr3}) at small distances. 
As we observe in Fig.~\ref{fig1}, $P(r)$ starts to decrease 
rapidly around the surface of star. 
In the numerical simulation of Fig.~\ref{fig1}, 
the star radius is $R_{\ast} \simeq 9.3$~km for $\beta_3=0$. 
In the presence of negative coupling $\beta_3$, the pressure 
decreases more slowly with increasing $r$, see case (i) 
of Fig.~\ref{fig1}. In case (i), we have chosen a smaller value 
of the central pressure relative to that in GR, 
but the smaller decreasing rate of $P(r)$ in the former 
leads to the larger radius, $R_{\ast} \simeq 12.3$~km.
The case (ii) in Fig.~\ref{fig1} corresponds to a positive value 
of $\beta_3$ with a larger central pressure compared 
to the GR case. The decreasing rate of $P(r)$ in case (ii) 
is faster than that in GR, so the resulting radius 
is smaller, $R_{\ast} \simeq 8.3$~km.

Similarly, the density $\rho(r)$ also decreases as a function 
of $r$. 
The central density $\rho_c$ in case (i) is smaller than 
that in GR, while the radius $R_{\ast}$ is larger. 
Since the density $\rho(r)$ in case (i) decreases more slowly 
relative to the case of GR, the former catches up with the 
latter at an intermediate distance ($r \simeq 6$~km). 
The $r$-derivative of the mass function $M(r)$ can be 
generally written in the form 
\be
M'(r)=4\pi \rho(r) r^2 [1+\tilde{\cal F}(r) ]\,,
\label{dM}
\ee
where $\tilde{\cal F}(r)$ is a function of $r$ containing 
the dependence of $\beta_3$. 
As we estimated in Eq.~(\ref{Mass}), the functions
$\rho(r)$ and $\tilde{\cal F}(r)$ around $r=0$ reduce 
to the constants $\rho_c$ and ${\cal F}$, respectively.
When we integrate Eq.~(\ref{dM}) with respect to $r$, the first 
term on the r.h.s. gives rise to a contribution to $M_{\ast}$ which 
is roughly proportional to  $(4\pi/3)\rho_c R_{\ast}^3$.
The increase of $R_*$ induced by the negative coupling 
$\beta_3$ leads to a larger contribution to $M_{\ast}$ 
relative to the decrease of $\rho_c$. 
In case (i) the mass contribution arising from 
the integration of the term $4\pi \rho(r) r^2$ in Eq.~(\ref{dM}) 
is $M_{{\ast}1}=2.87M_{\odot}$, which is larger than  
the value $M_{{\ast}1}=1.67M_{\odot}$ of GR in Fig.~\ref{fig1}. 
The ratio of $R_{\ast}^3$ between the case (i) and GR is 
given by $(12.3/9.3)^3=2.31$. 
This increase is slightly compensated by the smaller 
density in the central region with the decrease 
about 25 \%, so the resulting ratio of $M_{{\ast}1}$ between 
the two cases becomes $2.87/1.67=1.72<2.31$.

For $\beta_3<0$ the function $\tilde{{\cal F}}(r)$
in Eq.~(\ref{dM}) is negative around $r=0$, so 
the negative coupling works to 
reduce the mass term $M_{{\ast}1}$. 
In case (i) of Fig.~\ref{fig1}, the mass $M_{{\ast}2}$ arising 
from the numerical integration of $4\pi \rho(r) r^2 
\tilde{{\cal F}}(r)$ is found to be $M_{{\ast}2}\simeq 
-0.29M_{{\ast}1}$, so the total mass 
$M_{\ast}=M_{{\ast}1}+M_{{\ast}2}$ can be estimated as 
$M_{\ast} \simeq 0.71 M_{{\ast}1} \simeq 2.03M_{\odot}$.
The mass function $M(r)$ 
in case (i) is smaller than that in GR except for the distance 
$r$ around the surface of star.
However, the increase of $M(r)$ in case (i) continuously 
occurs up to the radius $R_{\ast}$ larger than that in GR, 
so the resulting mass $M_{\ast}$ in the former is larger. 
Thus, the main reason for the increase of $M_{\ast}$ 
comes from the increase of $R_{\ast}$ induced 
by the negative coupling.
 
For $\beta_3>0$, the radius $R_{\ast}$ gets smaller 
compared to the value in GR, see case (ii) of Fig.~\ref{fig1}.
Since the function $\tilde{{\cal F}}(r)$ in Eq.~(\ref{dM}) 
is positive, the mass function $M(r)$ is larger than that 
in GR at small distances. 
However, the increase of $M(r)$ stops at a smaller 
radius $R_{\ast}$, which results in a smaller mass 
$M_{\ast}$. Hence the positive coupling $\beta_3$ 
generally leads to the decrease of mass
$M_{\ast}$ relative to the GR case.

\begin{figure}[h]
\begin{center}
\includegraphics[height=3.3in,width=3.5in]{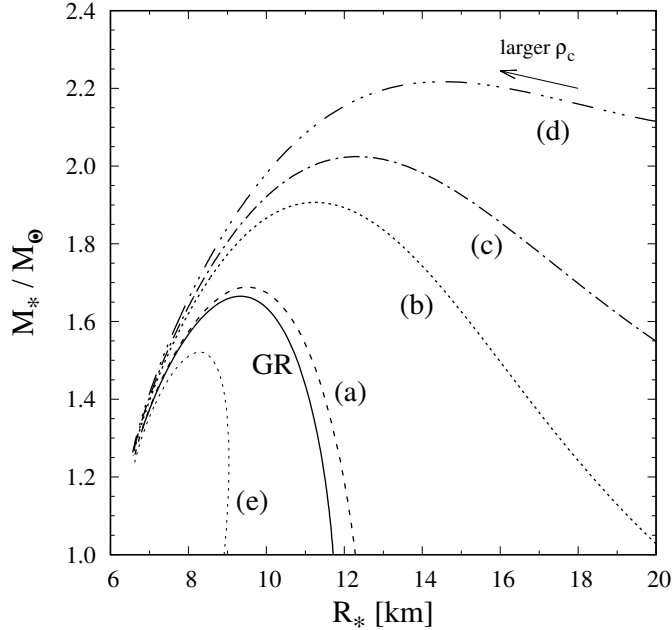}
\end{center}
\caption{\label{fig2}
Mass-radius relations in cubic Galileons 
for the polytropic EOS (\ref{eos}) with 
$\Gamma=2.34$, $K=0.0130$. 
We choose the boundary conditions 
(\ref{G3ini1})-(\ref{G3ini4}) 
at the distance $r=10^{-3}r_0$.
Each curve corresponds to 
(a) $\tilde{\beta}_3=-1$, $\bar{a}_0=1.0$, 
(b) $\tilde{\beta}_3=-1$, $\bar{a}_0=2.0$, 
(c) $\tilde{\beta}_3=-1$, $\bar{a}_0=2.2$, 
(d) $\tilde{\beta}_3=-1$, $\bar{a}_0=2.4$, 
(e) $\tilde{\beta}_3=+1$, $\bar{a}_0=2.0$, and 
the GR case $\tilde{\beta}_3=0$, $\bar{a}_0=0$.
With increasing $\rho_c$, the values of $M_*$ and $R_*$ 
shift to the direction shown as the arrow inside the figure. 
}
\end{figure}

In Fig.~\ref{fig2}, we plot the 
mass-radius relation for the polytropic EOS (\ref{eos}) 
with $K=0.0130$ and $\Gamma=2.34$. 
The central density is chosen to be in the range  
$y_c=\rho_c/\tilde{\rho}_0 \le 200$.
In this case, the maximum ADM mass $M_{\ast}$ 
in GR is given by 
$M_{\rm max}=1.67M_{\odot}$ with the radius 
$R_{\ast}=9.3$ km and the central density 
$\rho_c=3.5 \times 10^{15}$ g\,$\cdot$\,cm$^{-3}$ (plotted 
as the GR case in Fig.~\ref{fig1}). 
For increasing negative coupling $|\beta_3|$, the maximum mass 
gets larger. This effect tends to be significant for 
$|\tilde{\beta}_3| \bar{a}_0^2$ 
exceeding the order of 1. The maximum mass reached for 
$\tilde{\beta}_3=-1$ and $\bar{a}_0=2.2$ (case (c) in 
Fig.~\ref{fig2}) is $M_{\rm max}=2.03M_{\odot}$ with the radius 
$R_{\ast}=12.3$~km and the central density 
$\rho_c=2.1 \times 10^{15}$~g\,$\cdot$\,cm$^{-3}$ 
(plotted as case (i) of Fig.~\ref{fig1}). 
Even though $\rho_c$ is smaller than that in GR, the larger 
radius $R_{\ast}$ leads to the maximum mass $M_{\rm max}$ 
which is about $2.03/1.67=1.22$ times as large as that in GR. 

{}From Eq.~(\ref{bebo3}), there is the constraint
$|\bar{\beta}_3|\bar{a}_0^2<4\sqrt{2\pi/
[3(2+3w_c\bar{a}_0^2)]}$ 
for $n=1$. If $\tilde{\beta}_3=-1$, $w_c=0.247$, $\rho_c/\rho_0=12.8$, 
this bound translates to $\bar{a}_0<2.7$. 
For increasing $\bar{a}_0$, the resulting mass of star 
tends to be larger. 
In case (d) shown in Fig.~\ref{fig2} 
($\tilde{\beta}_3=-1$, $\bar{a}_0=2.4$), 
the maximum mass for the radius 
$R_*<20$~km is given by $M_{\rm max}=2.22M_{\odot}$. 
For $2.5 \lesssim \bar{a}_0<2.7$, $M_{\ast}$ changes to a continuously growing
function with respect to $R_{\ast}$. 
This property may be understood by using Eq.~(\ref{G3ini4}) 
for  $|\bar{\beta}_3|\bar{a}_0^2$ close to the upper 
bound (\ref{bebo3}). 
In this case, the star radius can be crudely estimated as 
\be
R_* \approx r_0 \sqrt{\frac{\rho_0}{2\pi(1+w_c)\rho_c}}\,.
\label{Rses}
\ee
In the regime $w_c \ll 1$, the radius has the dependence 
$R_* \propto \rho_c^{-1/2}$, so it increases for decreasing $\rho_c$. 
The quantity $\rho_c R_*^3$ also increases for smaller $\rho_c$, as 
$\rho_cR_*^3 \propto \rho_c^{-1/2} \propto R_*$. 
While the negative coupling $\bar{\beta}_3$ suppresses the growth of 
$M(r)$ around $r=0$, this is compensated by the increase of $R_*$ 
in the region of small $\rho_c$.
Hence, for $|\bar{\beta}_3|\bar{a}_0^2$ close to the upper 
bound (\ref{bebo3}), the mass $M_{\ast}$ continuously grows 
with the increase of $R_{\ast}$.
Unless $\bar{a}_0$ is very close to the upper limit $2.7$, 
the maximum mass $M_{\rm max}$ does not exceed 
$3M_{\odot}$ for $R_{\ast}<20$ km with 
the model parameters used in Fig.~\ref{fig2}.

If the quantity $|\bar{\beta}_3|\bar{a}_0^2$ exceeds 
the upper limit set by Eq.~(\ref{bebo3}), the mass function $M(r)$ is negative around the center of star.  
Indeed, we numerically confirmed that the mass function 
enters the region $M(r)<0$ around $r=0$ and then $M(r)$ becomes positive at the distance away from the center.
We regard that this situation is unphysical. 

\begin{figure}[h]
\begin{center}
\includegraphics[height=3.4in,width=3.7in]{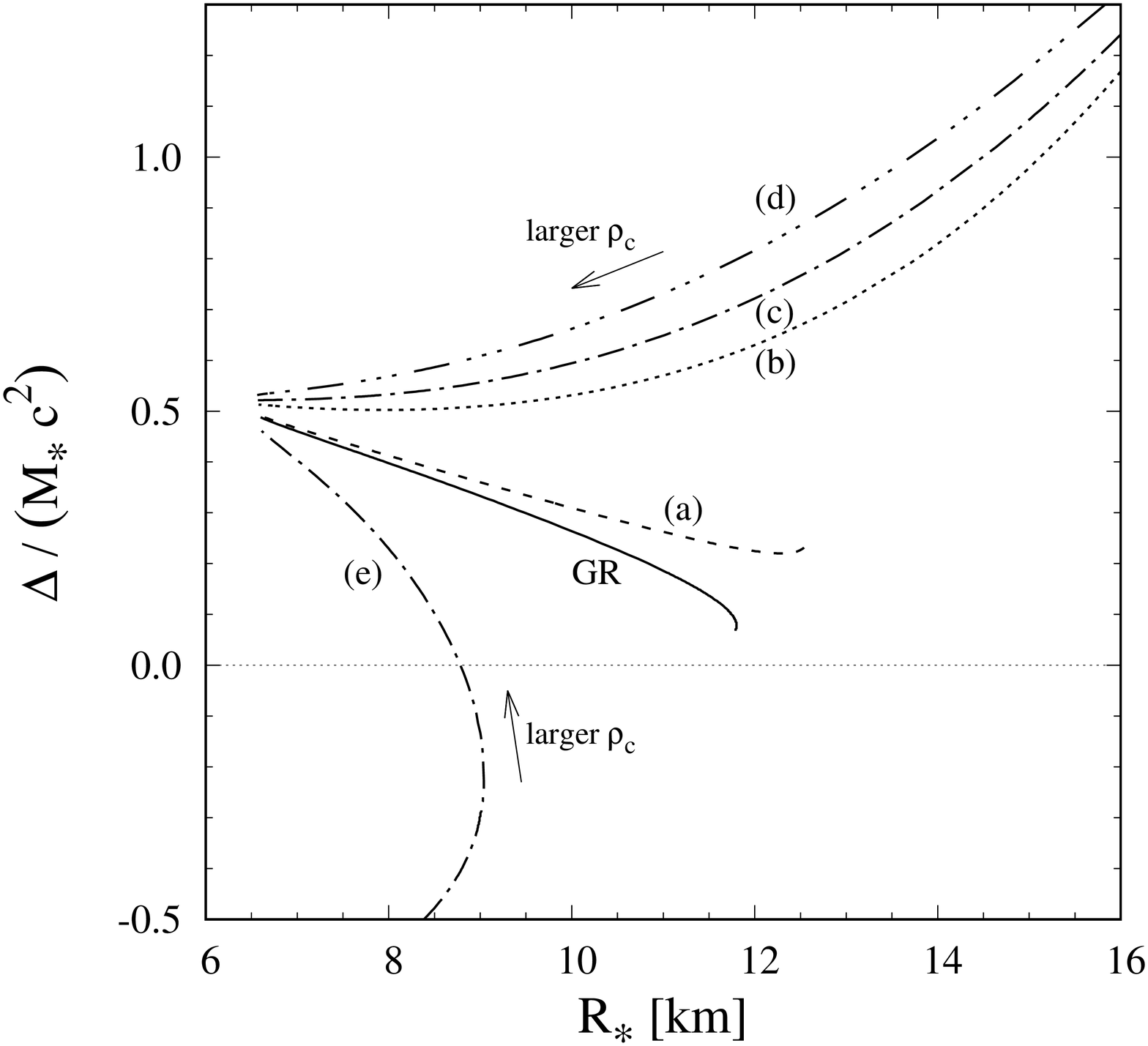}
\end{center}
\caption{\label{fig3}
The gravitational binding energy $\Delta$ normalized 
by $M_{\ast} c^2$ versus the radius $R_{\ast}$ 
in cubic Galileons for the polytropic EOS (\ref{eos}) 
with $K=0.0130$ and $\Gamma=2.34$ in the region 
of the central density $3 \le y_c \le 200$.
Each curve corresponds to the cases plotted in Fig.~\ref{fig2}. 
With increasing $\rho_c$, the values of $\Delta$ and 
$R_{\ast}$ shift to the direction shown as the arrow inside the figure.}
\end{figure}

\begin{figure}[h]
\begin{center}
\includegraphics[height=3.1in,width=3.3in]{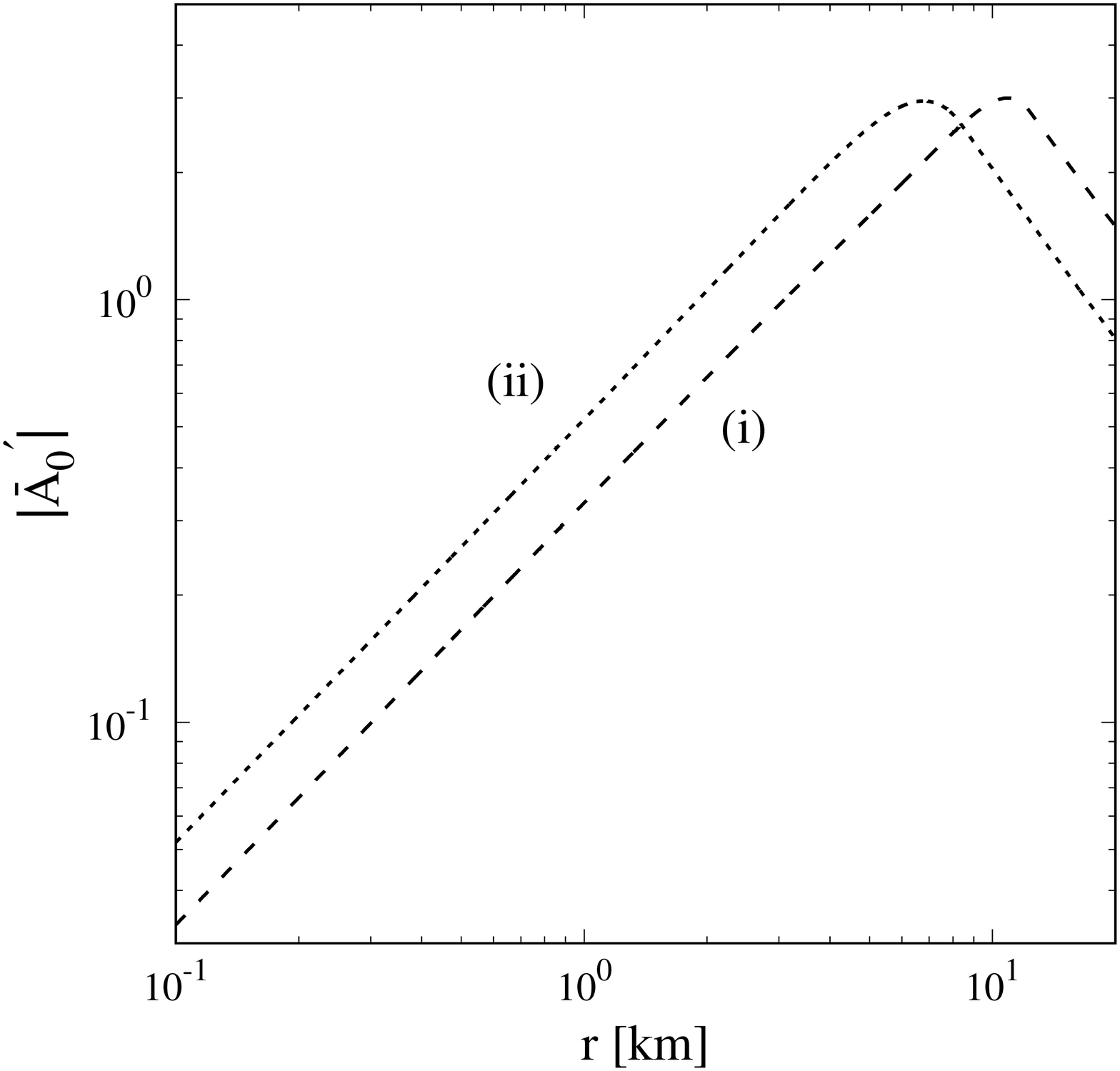}
\includegraphics[height=3.1in,width=3.3in]{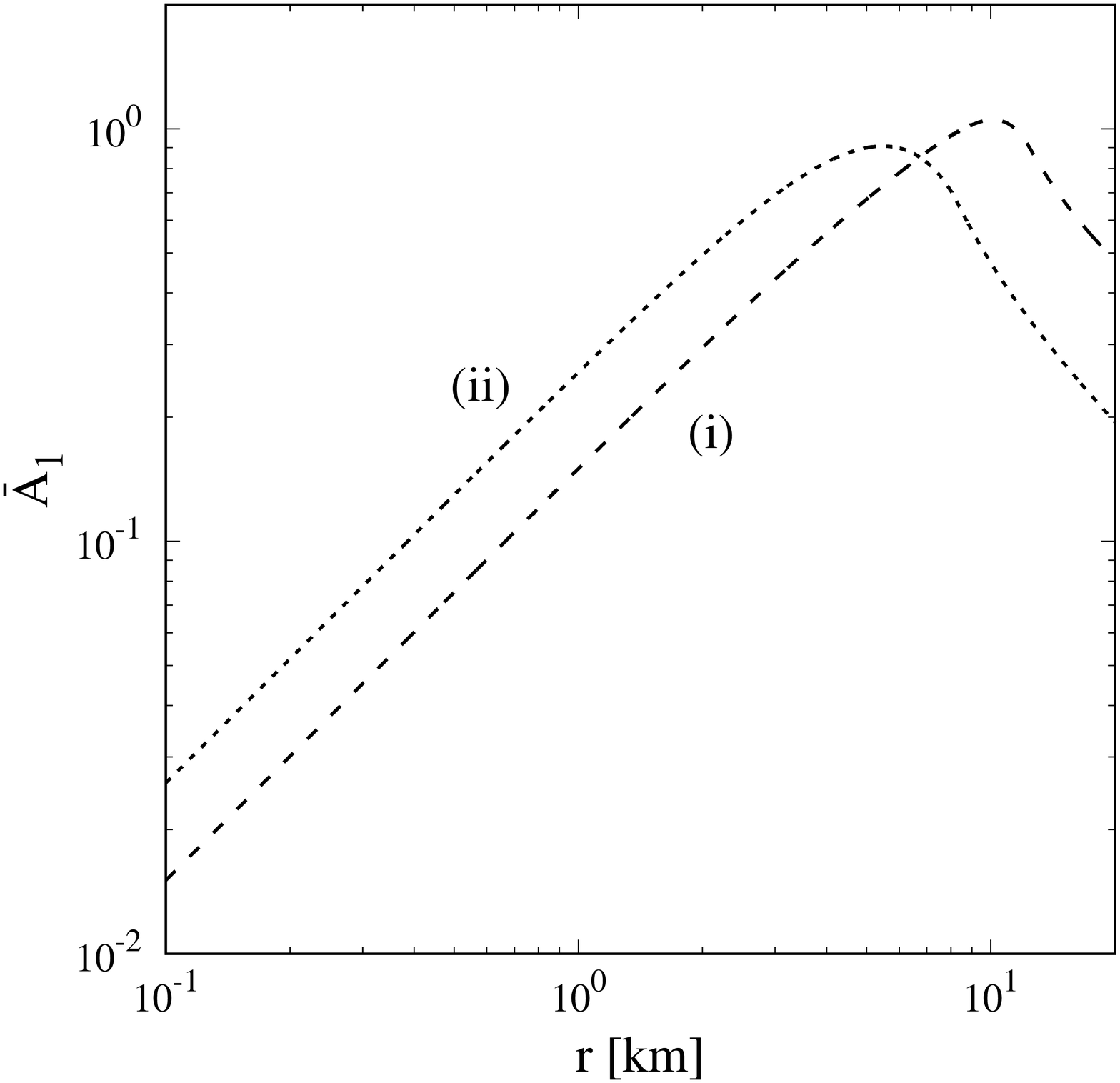}
\end{center}
\caption{\label{fig4}
Numerical solutions to the derivative 
$|\bar{A}'_0|=\sqrt{8\pi G}\,r_0|A_0'|$ (left) and 
the longitudinal mode $\bar{A}_1=\sqrt{8\pi G}A_1$ 
(right) in cubic Galileons for the polytropic 
(\ref{eos}) with $K=0.0130$ and $\Gamma=2.34$. 
The cases (i) and (ii) correspond to the same model parameters 
and boundary conditions as those used in Fig.~\ref{fig1}.
}
\end{figure}

As we see in case (e) of Fig.~\ref{fig2}, 
the positive coupling $\beta_3$ 
leads to smaller $M_{*}$ and $R_{*}$ than those in GR.
In Fig.~\ref{fig3}, we plot the quantity 
$\Delta/(M_{\ast}c^2)=M_p/M_{\ast}-1$ versus the radius 
$R_{\ast}$ for the same model parameters as those used in Fig.~\ref{fig2}, 
where $M_p$ is the proper mass defined by Eq.~(\ref{Mp}). 
For $\beta_3<0$, the binding energy $\Delta$ is always 
positive, so the star is gravitationally bound.
For $\beta_3>0$, the star tends to be dynamically 
unstable in the region of small $\rho_c$. 
The configuration of maximum mass $M_{\ast}=1.52M_{\odot}$ in case (e) of Fig.~\ref{fig2} 
($\tilde{\beta}_3=+1$ and $\bar{a}_0=2.0$), which corresponds to the central density 
$\rho_c=4.3 \times 10^{15}$ g\,$\cdot$\,cm$^{-3}$ 
and the radius $R_{\ast}=8.3$ km, 
leads to a positive binding energy, but
the sign of $\Delta$ changes to negative 
for $\rho_c<2.6 \times 10^{15}$ g\,$\cdot$\,cm$^{-3}$.

For the ranges of $\rho_c$ smaller than those plotted 
as the cases (b), (c), (d) of Fig.~\ref{fig3}, 
we numerically find that there is a maximum value of 
$\Delta/(M_*c^2)$ and then the binding energy gets smaller for decreasing $\rho_c$ further. 
In the intermediate regime where $\Delta/(M_*c^2)$ 
deceases with the increase of $\rho_c$, there is the 
``repulsive'' gravity effect  induced by the negative coupling 
$\beta_3$. 
The pressure increased by the negative coupling $\beta_3$ 
can support the star with 
a stronger gravitational force. 
In other words, the increased binding energy in the intermediate 
regime of $\rho_c$ is compatible with the large effective 
pressure induced by $\beta_3$.

The above discussion shows that not only the sign and 
the strength of coupling $\beta_3$ but also the amplitude 
of $A_0$ plays an important role for increasing
the mass and radius of star. 
Around $r=0$, the temporal component is 
given by Eq.~(\ref{G3ini3}), so the derivative $|A_0'|$ grows 
in proportion to $r$. The longitudinal mode $A_1$ has the same 
$r$-dependence as $|A_0'|$ around $r=0$, see Eq.~(\ref{A1ho}).
For increasing $|a_0|$, the amplitude of $A_1$ also tends to be larger.
In Fig.~\ref{fig4}, we plot $|A_0'|$ and $A_1$ versus $r$ 
for the cases (i) and (ii) shown in Fig.~\ref{fig1}. 
In both cases, $|A_0'|$ and $A_1$ increase in proportion to $r$ 
up to the distance close to the surface of star. 
Outside the body ($r>R_{\ast}$), the behavior of 
vector field is similar to the vacuum solution around
the static and 
spherically symmetric BHs derived 
in Refs.~\cite{HKMT,HKMT2}.
Namely, both $|A_0'|$ and $A_1$ decrease as 
$\propto 1/r^2$ for $r \gg R_{\ast}$. 
As in Refs.~\cite{HKMT,HKMT2}, the coupling 
$\beta_3$ induces some difference between the two metric components $f$ and $h$ around
the surface of star, but the difference becomes negligible in the 
limit that $r \gg R_{\ast}$.

\section{Quartic couplings}
\label{sec5}

In this section, we study the effect of quartic derivative couplings $G_4(X)$ on the configuration 
of relativistic stars. 
We consider the  power-law coupling model given by 
\be
G_4=\frac{1}{16\pi G}
+\beta_4 X^n\,,
\label{G4}
\ee
with $G_2=G_3=G_5=G_6=0$ and $g_5=0$, 
where $\beta_4$ is a constant and $n$ is a positive integer.
In Ref.~\cite{Tasinato}, the authors discussed the relativistic 
star solutions for the specific case $n=1$. 
Now, we investigate the models of general power $n$ 
including the quartic vector Galileon ($n=2$).
{}From Eq.~(\ref{be5}), the longitudinal mode obeys 
\be
\beta_{4}A_{1}
\left( A_0^{2} -fh A_{1}^{2}  \right) ^{n-2}
[A_1^2 fh \{ (1+h-2nh)f+(1-2n)rf'h \}+
A_0^2 \{ f(h-1)+(2n-1)rf'h \}-4r(n-1)A_0A_0'fh]=0\,.
\label{be5G4}
\ee
This gives rise to the two branches characterized by $A_1=0$ or $A_1 \neq 0$. 
For the latter branch, our numerical analysis shows that the solutions are qualitatively similar to those of cubic derivative couplings discussed 
in Sec.~\ref{sec4}. 
Hence we will focus on the other branch
\be
A_1=0\,,
\ee
in the rest of this section.

\subsection{Analytic solutions around the center of star}

Let us first derive analytic solutions to $f,h,A_0, P$ 
by using the expansions (\ref{fh}) and (\ref{fhA0}) 
around $r=0$. {}From the continuity equation (\ref{mattereq}), 
we obtain the relation same as Eq.~(\ref{p2}) among the 
coefficients $p_2$ and $f_2$.
Substituting $A_1=0$ and $A_1'=0$ into 
Eqs.~(\ref{be1})-(\ref{be4}), we obtain the iterative solutions 
\ba
&&f(r)=1+f_2r^2+{\cal O} (r^4) 
\label{G4ini1}\,,\\
&&h(r)=1-\frac{8\pi}{3[1-2^{1-n}(2n-1)\bar{\beta}_4\at^{2n}]}\frac{r^2}{r_c^2}+{\cal O} (r^4)\,,
\label{G4ini2}\\
&&A_0(r)=\frac{1}{\sqrt{8\pi G}}
\left[\at+\frac{2^{4-n}\pi n\bar{\beta}_4\at^{2n-1}}
{3 \{1-2^{1-n}(2n-1)\bar{\beta}_4\at^{2n}\}}\frac{r^2}{r_c^2}\right]
+{\cal O} (r^4)\,,
\label{G4ini3}\\
&&P(r)=p_c
-\frac{c^2\rho_c+p_c}{2}f_2r^2
+{\cal O}(r^4)\,,
\label{G4ini4}
\ea
where the definitions of $\bar{a}_0, r_c, w_c$ are the 
same as those given in Eq.~(\ref{wcdef}), and 
\ba
& &
\bar{\beta}_4=\frac{\beta_4}{(8\pi G)^{n-1}}\,,\\
& &
f_2=
\frac{4\pi \left[1+3w_c+2^{1-n}\left\{1-3(2n-1)w_c\right\}\bar{\beta}_4\at^{2n}
-2^{5-2n}n^2\bar{\beta}_4^2\at^{4n-2}\right]}
{3[1-2^{1-n}(2n-1)\bar{\beta}_4\at^{2n}]^2r_c^2}\,.
\label{f2ex}
\ea

Without loss of generality, we assume that $\bar{a}_0>0$ 
in the following discussion.
On using Eq.~(\ref{f2ex}), the condition (\ref{matcon}) 
translates to 
\be
{\cal F}_{-}<\bar{\beta}_4 \at^{2n-2}
<{\cal F}_{+}\,,
\label{Fmp}
\ee
where ${\cal F}_{\pm}$ are defined by 
\be
{\cal F}_{\pm} \equiv \frac{2^{n-5}}{n^2}\left[1-3(2n-1)w_c
\pm\sqrt{\left\{1-3(2n-1)w_c\right\}^2
+32n^2(1+3w_c)\at^{-2}}\right]\,.
\ee
{}From Eqs.~(\ref{hr}) and (\ref{G4ini2}), 
the mass function around $r=0$ is given by 
\be
M(r)=
\frac{4\pi \rho_c r^3}{3[1-2^{1-n}(2n-1)\bar{\beta}_4\at^{2n}]}
+{\cal O}(r^5)\,.
\ee
To ensure that $M(r)>0$ around the center of star,
we require the condition 
\be
\bar{\beta}_4\bar{a}_0^{2n} <
\frac{2^{n-1}}{2n-1}\,,
\label{G4con}
\ee
which is automatically satisfied for $\bar{\beta}_4<0$. 
If $\bar{\beta}_4>0$, the upper limit corresponding 
to Eq.~(\ref{G4con}) leads to the divergence of the 
quantity $f_2$ in Eq.~(\ref{f2ex}), so the condition 
$\bar{\beta}_4\bar{a}_0^{2n-2}<{\cal F}_+$ 
gives the tighter bound than Eq.~(\ref{G4con}). 

The coupling $\bar{\beta}_4$ affects the decreasing rate 
of the pressure $P(r)$ through the function $f_2$, whose 
value in GR is given by 
$f_{2}^{{\rm GR}}=4\pi(1+3w_c)/(3r_c^2)$. 
The difference between $f_2$ and 
$f_{2}^{{\rm GR}}$ is
\be
f_2-f_{2}^{{\rm GR}}=
\frac{8\pi \bar{\beta}_4 \bar{a}_0^{2n}
[2^n \{ 4n-1+3w_c(2n-1) \}-2\bar{\beta}_4 
\bar{a}_0^{2n-2} \{8n^2+ \bar{a}_0^2 
(2n-1)^2(1+3w_c)\}]}
{3[2^n-2 \bar{\beta}_4 \bar{a}_0^{2n} (2n-1)]^2r_c^2}\,.
\label{f2di}
\ee

For $\bar{\beta}_4<0$, the r.h.s. of Eq.~(\ref{f2di}) is 
negative and hence $f_2<f_{2}^{{\rm GR}}$. 
If the solution (\ref{G4ini4}) is extrapolated up to the 
surface of star, it is expected that the radius $R_{\ast}$ 
is larger than that in GR due to the slower decrease 
of $P(r)$ toward 0. Since the amplitude of negative coupling 
is not constrained from the condition 
$M(r)>0$, the radius $R_{\ast}$ is not bounded from above.
This property is different from that in cubic power-law couplings 
where $R_{\ast}$ is constrained as Eq.~(\ref{Rstar}) from 
the condition $M(r)>0$.

If $\bar{\beta}_4>0$ and $|\bar{\beta}_4| \bar{a}_0^{2n-2} \ll 1$, then the first term in the square bracket of the numerator of Eq.~(\ref{f2di}) dominates over the second one, so that $f_2>f_{2}^{{\rm GR}}$. 
In this regime, the radius $R_{\ast}$ should be smaller than that 
in GR  due to the faster decrease of $P(r)$ toward 0.
For increasing $\bar{\beta}_4$ and $\bar{a}_0$,  
the function $f_2$ reaches a maximum and then 
it starts to decrease toward $0$ (which 
corresponds to $\bar{\beta}_4\bar{a}_0^{2n-2}
={\cal F}_{+}$).
As a function of $\bar{\beta}_4$, $f_2$ has
the maximum value 
\be
f_2^{\rm max}=\frac{\pi[32n^2(1+3w_c)
+\bar{a}_0^2 \{1+3(1-2n)w_c \}^2]}
{6n[4n-\bar{a}_0^2(2n-1)]r_c^2}\,,
\ee
at 
\be
\bar{\beta}_4=\frac{2^{n-1}
[2(2+3w_c)n-1-3w_c]}{16n^2-\bar{a}_0^2(2n-1)
[1+3(1-2n)w_c]}\bar{a}_0^{2-2n}\,.
\label{be4m}
\ee
The coupling (\ref{be4m}) is smaller than the upper limit 
$\bar{\beta}_4={\cal F}_{+}\bar{a}_0^{2-2n}$ determined by 
Eq.~(\ref{Fmp}). This gives the following bound 
\be
\bar{a}_0<\bar{a}_{\rm max} \equiv \sqrt{\frac{4n}{2n-1}}\,.
\label{a0limit}
\ee
The regime in which the condition $f_2<f_2^{\rm GR}$ 
is satisfied is given by 
\be
{\cal F}_c\bar{a}_0^{2-2n}
<\bar{\beta}_4<{\cal F}_{+}\bar{a}_0^{2-2n}\,,
\label{calFcon}
\ee
where 
\be
{\cal F}_c \equiv
\frac{2^{n-1}[2(2+3w_c)n-1-3w_c]}
{8n^2+\bar{a}_0^2(2n-1)^2(1+3w_c)}\,.
\ee
To realize the slower decrease of $P(r)$ around $r=0$ 
relative to the GR case, we need to choose the large 
value of $\bar{\beta}_4\bar{a}_0^{2n-2}$ close to 
${\cal F}_{+}$. For given $\bar{\beta}_4$ and $n>1$,
this amounts to choosing larger $\bar{a}_0$ close to 
the upper bound (\ref{a0limit}). 
Taking the limit $\bar{a}_0 \to \bar{a}_{\rm max}$, however, 
both ${\cal F}_c\bar{a}_0^{2-2n}$ and ${\cal F}_{+}\bar{a}_0^{2-2n}$ approach the same value 
$(2n-1)^{n-1}/(2^{n+1}n^n)$. 
In this limit, the parameter space consistent with 
Eq.~(\ref{calFcon}) disappears with the 
divergence of $f_2^{\rm max}$.
Even if we consider the value $\bar{a}_0=
\bar{a}_{\rm max}-\varepsilon$, where $\varepsilon$ is 
a small positive parameter, the expansions of 
${\cal F}_c\bar{a}_0^{2-2n}$ and ${\cal F}_{+}\bar{a}_0^{2-2n}$ in terms of $\varepsilon$ show that two 
terms are equivalent up to the order of $\varepsilon$.
Since the difference between ${\cal F}_c\bar{a}_0^{2-2n}$ and ${\cal F}_{+}\bar{a}_0^{2-2n}$ appears only at the 
order of $\varepsilon^2$, the parameter space consistent with 
Eq.~(\ref{calFcon}) is restricted to be very narrow. 
This discussion shows that, for $\bar{\beta}_4>0$, the function 
$f_2$ is in the range $f_2>f_2^{\rm GR}$ for most of the parameters under consideration, which should result in 
smaller $R_{\ast}$ compared to the GR case.

In the following, we will confirm the above analytic estimation 
by numerically solving Eqs.~(\ref{be1})-(\ref{be4}) and (\ref{mattereq}) with $A_1=0$.

\subsection{Numerical solutions}
\label{G4nu}

For the numerical computation, we focus on the case of 
quartic vector Galileons ($n=2$). The property of 
solutions in other power-law models ($n \neq 2$) 
are qualitatively similar to those discussed below.

\begin{figure}[h]
\begin{center}
\includegraphics[height=2.9in,width=3.1in]{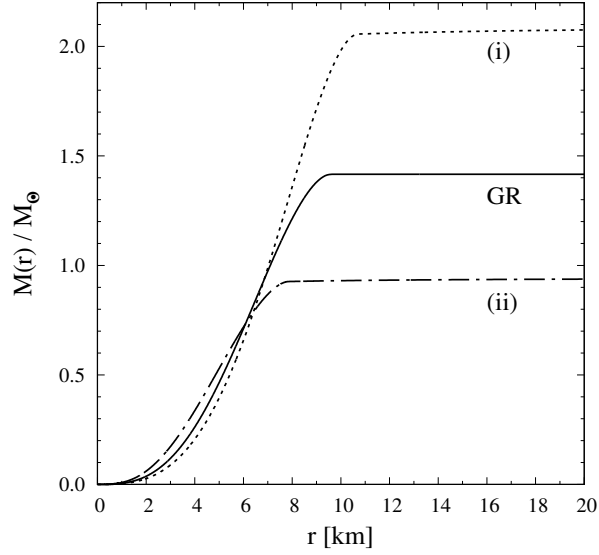}
\end{center}
\caption{\label{fig5}
The mass function $M(r)$ versus the distance $r$ in 
quartic Galileons $(n=2)$ for the polytropic EOS 
(\ref{eos}) with $\Gamma=2.34$, $K=0.010$, and 
the central density 
$\rho_c=2.3 \times 10^{15}$ g\,$\cdot$\,cm$^{-3}$.
The two curves at the top and bottom correspond to 
the model parameters
(i) $\bar{\beta}_4=-0.06$, $\bar{a}_0=1.5$, and 
(ii) $\bar{\beta}_4=0.06$, $\bar{a}_0=1.5$, while the 
solid curve corresponds to GR with 
$\bar{\beta}_4=0$, $\bar{a}_0=0$. }
\end{figure}

\begin{figure}[h]
\begin{center}
\includegraphics[height=3.1in,width=3.3in]{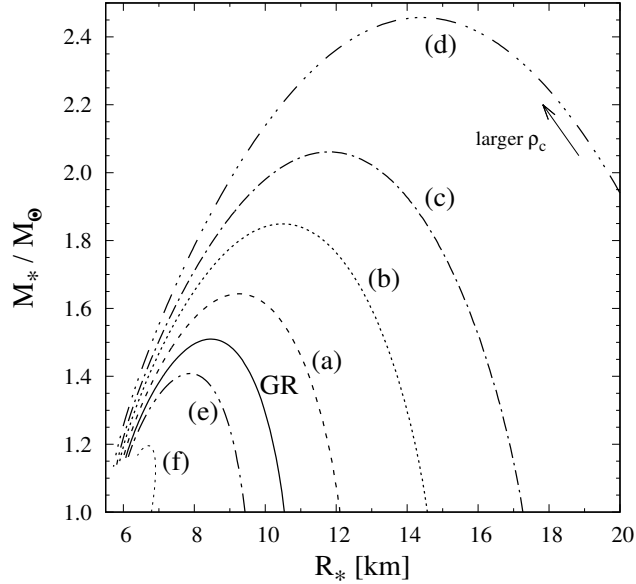}
\end{center}
\caption{\label{fig6}
Mass-radius relations in quartic Galileons  
for the polytropic EOS (\ref{eos}) with 
$\Gamma=2.34$, $K=0.01$. 
We use Eqs.~(\ref{G4ini1})-(\ref{G4ini4}) as the 
boundary conditions at the distance $r=10^{-3}r_0$. 
Each curve corresponds to 
(a) $\bar{\beta}_4=-0.1$, $\bar{a}_0=1.0$, 
(b) $\bar{\beta}_4=-0.1$, $\bar{a}_0=1.2$, 
(c) $\bar{\beta}_4=-0.1$, $\bar{a}_0=1.3$, 
(d) $\bar{\beta}_4=-0.1$, $\bar{a}_0=1.4$, 
(e) $\bar{\beta}_4=+0.1$, $\bar{a}_0=1.0$, and
(f) $\bar{\beta}_4=+0.1$, $\bar{a}_0=1.5$.
The GR case with $\bar{\beta}_4=0$ and 
$\bar{a}_0=1.0$ is plotted as the solid line.
With increasing $\rho_c$, the values of $M_{\ast}$ 
and $R_{\ast}$ shift to the direction shown as 
the arrow inside the figure. 
}	
\end{figure}

\begin{figure}[h]
\begin{center}
\includegraphics[height=3.3in,width=3.5in]{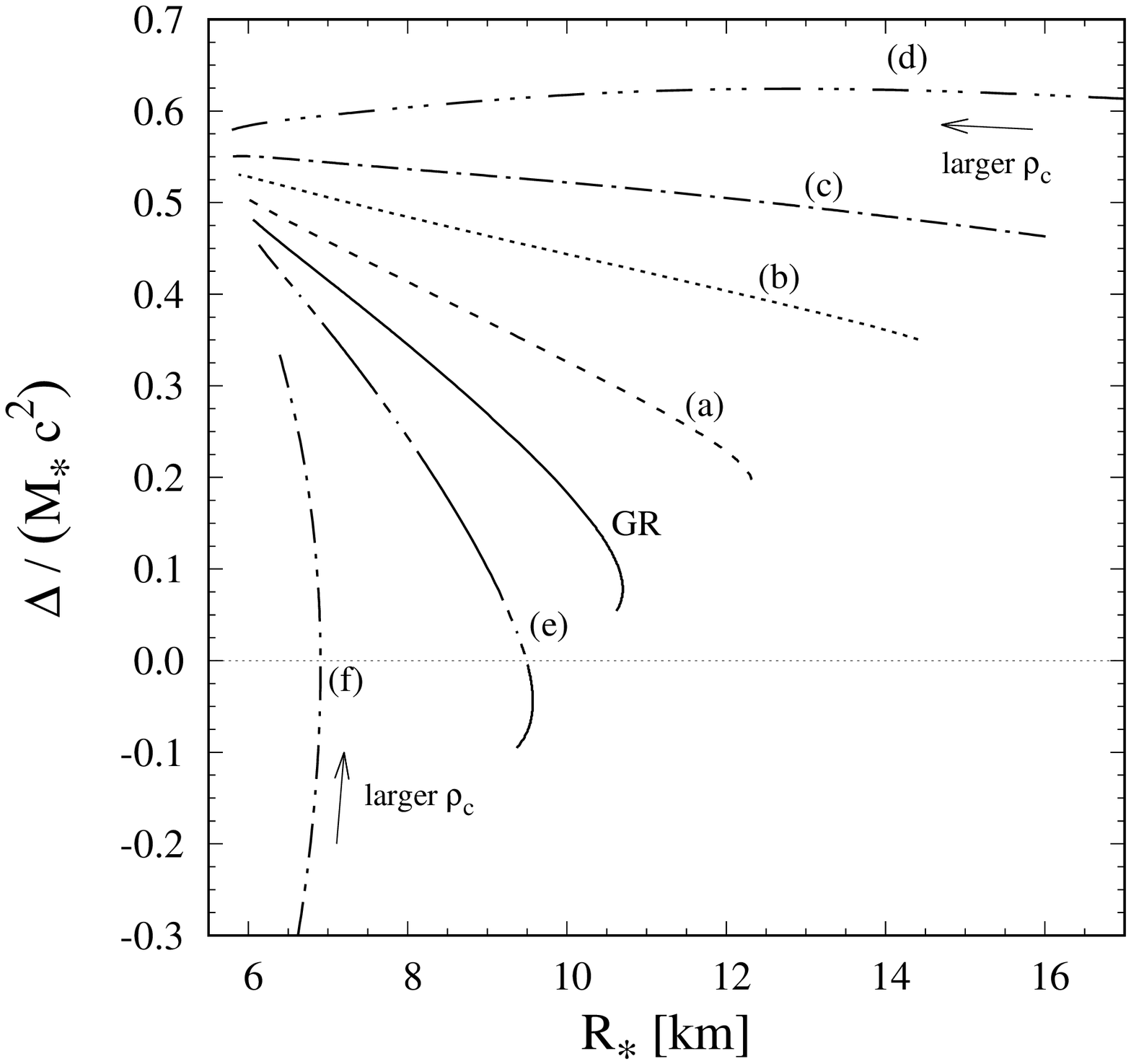}
\end{center}
\caption{\label{fig7}
The binding energy $\Delta$ normalized by $M_{\ast}c^2$ 
versus the radius $R_{\ast}$ in quartic Galileons 
for the polytropic EOS with $\Gamma=2.34$ and 
$K=0.01$ in the range $3 \le y_c \le 200$.
Each curve corresponds to the cases plotted in Fig.~\ref{fig6}.}
\end{figure}

In Fig.~\ref{fig5}, we plot the mass function $M(r)$ versus $r$ 
for several different values of $\bar{\beta}_4$ and $\bar{a}_0$ 
with the same central density $\rho_c$. 
We employ the polytropic EOS (\ref{eos}) 
with $\Gamma=2.34$ and $K=0.01$. 
The mass $M_{\ast}$ and the radius $R_{\ast}$ of 
star can be identified by the point at which $M(r)$ 
stops increasing, e.g., $M_{\ast} \simeq 1.4M_{\odot}$ 
and $R_{\ast} \simeq 9.5$ km in GR ($\bar{\beta}_4=0$).
As we analytically estimated above, the value of $M(r)$ 
for $\bar{\beta}_4<0$ is smaller than that in GR 
at small distances. However, as we see in case (i) of 
Fig.~\ref{fig5}, the mass function in the former 
catches up with that in the latter at an intermediate 
distance inside the star, so the resulting mass 
$M_*$ gets larger. Moreover, we have numerically 
confirmed that the negative coupling $\bar{\beta}_4$ leads to 
a slower decrease of the pressure $P(r)$ up to the star surface 
relative to the case $\bar{\beta}_4=0$, 
which results in a greater radius $R_{\ast}$.
The case (i) in Fig.~\ref{fig5} shows that both $R_*$ 
and $M_{\ast}$ are larger than those in GR. 
When $\bar{\beta}_4>0$, the mass function $M(r)$ at small 
distances is larger than that for $\bar{\beta}_4=0$.
This property can be seen in case (ii) of Fig.~\ref{fig5}, 
but the increase of $M(r)$ stops 
at a smaller radius $R_{\ast}$ because of 
a faster decrease of $P(r)$. 
Hence the mass $M_{\ast}$ in case (ii) is smaller 
than that in GR.
 
In Fig.~\ref{fig6}, we show the mass-radius 
relation for the polytropic EOS with $\Gamma=2.34$ and 
$K=0.01$ in the presence of quartic Galileon couplings 
$\bar{\beta}_4=-0.1$ or $\bar{\beta}_4=0.1$.
Compared to GR, the negative coupling 
$\bar{\beta}_4$ leads to larger values of $M_{\ast}$ 
and $R_{\ast}$. For this EOS, the maximum value of $M_*$ 
in GR is given by $M_{\rm max}=1.51M_{\odot}$ 
with the central density 
$\rho_c=4.1 \times 10^{15}$ g\,$\cdot$\,cm$^{-3}$ 
and the radius $R_{\ast}=8.48$ km. 
In the presence of negative $\bar{\beta}_4$, the larger 
maximum mass can be realized 
with the smaller central density. 
In case (c) plotted in Fig.~\ref{fig6}, 
which corresponds to $\bar{\beta}_4=-0.1$ and 
$\bar{a}_0=1.3$, the maximum 
mass $M_{\rm max}=2.06M_{\odot}$ with the radius 
$R_{\ast}=11.8$ km is reached at the density 
$\rho_c=1.6 \times 10^{15}$~g\,$\cdot$\,cm$^{-3}$. 
If we increase either 
$|\bar{\beta}_4|$ or $\bar{a}_0$ further, $M_{\rm max}$ 
becomes larger. Indeed, the condition $M(r)>0$ around $r=0$ does 
not restrict the amplitude of negative coupling 
$\bar{\beta}_4$, so the mass $M_{\ast}$ can be even 
larger than $3M_{\odot}$ for $\bar{\beta}_4$ close to the lower limit determined by the condition 
$\bar{\beta}_4 \bar{a}_0^{2}={\cal F}_{-}$.

In cases (e) and (f) depicted in Fig.~\ref{fig6}, which 
correspond to $\bar{\beta}_4>0$, the mass $M_{\ast}$ 
and the radius $R_{\ast}$ are smaller than those in GR,
independent of the detail of EOSs.
In these cases the condition $f_2>f_2^{\rm GR}$ 
is satisfied, so the faster decrease of $P(r)$ leads to 
the smaller radius $R_{\ast}$ compared to that in GR. 
As shown in Fig.~\ref{fig5}, the mass function $M(r)$ 
is larger than that for $\bar{\beta}_4=0$ in the central 
region of star, but the 
decrease of $R_{\ast}$ induced by positive $\bar{\beta}_4$ 
overwhelms this effect to end up with smaller $M_{\ast}$. 
We recall that there exists the restricted parameter range 
(\ref{calFcon}) in which the condition $f_2<f_2^{\rm GR}$ 
can be satisfied for $\bar{\beta}_4>0$.
When $n=2$, $\bar{\beta}_4=0.1$, and $w_c=0.4$, 
for example, the bound (\ref{calFcon}) translates to 
$1.600746<\bar{a}_0<1.600816$, whose parameter 
space is very narrow. Moreover, we find that the solutions 
in such a narrow parameter region are prone to 
numerical instabilities. 
Thus, the positive coupling $\bar{\beta}_4$ generally 
leads to the suppression of $M_{\ast}$ and $R_{\ast}$
in most of the parameter space 
with stable solutions.

The instability of star for large positive values of 
$\bar{\beta}_4\bar{a}_0^{2}$ close to ${\cal F}_{+}$
can be also confirmed by computing the binding energy 
$\Delta$ defined by Eq.~(\ref{Delta}). 
In Fig.~\ref{fig7}, we show $\Delta/(M_{\ast}c^2)$ versus 
the radius $R_{\ast}$ for $\Gamma=2.34$ and 
$K=0.01$ with several different 
values of $\bar{\beta}_4$ and $\bar{a}_0$. 
When $\bar{\beta}_4<0$, the binding energy is always 
positive, so the star is gravitationally bound.
If $\bar{\beta}_4>0$, $\Delta$ can be negative in the 
region of small central density $\rho_c$. 
In cases (e) and (f) shown in Fig.~\ref{fig7}, $\Delta$ is 
negative for $\rho_c<1.2\times 10^{15}$~g\,$\cdot$\,cm$^{-3}$ 
and $\rho_c<8.2\times 10^{15}$~g\,$\cdot$\,cm$^{-3}$, respectively, 
so that the region of instability tends to be larger 
for $\bar{\beta}_4\bar{a}_0^{2}$ approaching the 
upper limit ${\cal F}_{+}$. 
Thus, for $\bar{\beta}_4>0$, it is difficult to realize 
the stable configuration of star with 
$M_{\ast}$ and $R_{\ast}$
larger than those in GR.

\section{Intrinsic vector-mode couplings}
\label{sec6}

Finally, we investigate the relativistic star solutions 
in the presence of intrinsic vector-mode 
couplings given by 
\be
G_2=-2g_4(X)F\,,\qquad g_5=g_5(X)\,,\qquad
G_6=G_6(X)\,,\qquad 
\label{invec}
\ee
with $G_4=1/(16\pi G)$, 
where $g_4(X)$ is a function of $X$, and 
$F=hA_0'^2/(2f)$ on the background (\ref{metric}).
{}From Eq.~(\ref{be5}), it follows that 
\be
A_0'^2 \left[ \{ r^2 g_{4,X}+(3h-1)G_{6,X}\}A_1 
-2rg_5+2hr g_{5,X}A_1^2-G_{6,XX}h^2 A_1^3 
\right]=0\,.
\label{A1eqv}
\ee
We can write Eq.~(\ref{be4}) in the following form 
\be
\alpha_1 A_0''+\alpha_2A_0'+\alpha_3A_0'^2=0\,,
\label{A0dev}
\ee
where $\alpha_{1,2,3}$ are functions containing 
$A_0,A_1,A_1',f,h,f',h'$ and $g_4,g_5,G_6$ 
as well as their $X$-derivatives.  
The explicit expression of the coefficient $\alpha_1$ 
is given by 
\be
\alpha_1=(2g_4-1)r^2+4hr g_5 A_1
-2h^2A_1^2G_{6,X}+2(h-1)G_6\,.
\ee

{}From Eq.~(\ref{A1eqv}), there is a branch 
characterized by $A_0'(r)=0$, that is
\be
A_0(r)={\rm constant}\,,
\label{A0const}
\ee
which is consistent with Eq.~(\ref{A0dev}). 

There exist other branches where the terms in the square 
bracket of Eq.~(\ref{A1eqv}) vanish, which can give 
rise to a nonvanishing longitudinal component $A_1$.
Even in such cases, the temporal vector component 
needs to obey the regular boundary condition 
$A_0'(0)=0$ at the center of star. 
Then, we obtain $\alpha_1 A_0''(0)=0$ from Eq.~(\ref{A0dev}), 
so that $A_0''(0)=0$ 
for $\alpha_1 \neq 0$. 
This means that, when we integrate Eq.~(\ref{A0dev}) from 
$r=0$ with the boundary condition $A_0'(0)=0$, 
the derivative $A_0'(r)$ remains to be 0 
for arbitrary $r$. Then, provided that $\alpha_1 \neq 0$, 
we end up with the solution (\ref{A0const}) 
even for the branches other than $A_0'(r)=0$ 
in Eq.~(\ref{A1eqv}). 
Substituting the solution $A_0'(r)=0$ into Eqs.~(\ref{be1}) 
and (\ref{be2}), it follows that 
\ba
& &
\frac{h'}{r}+\frac{h-1}{r^2}=-\frac{8\pi G \rho}{c^2}\,,
\label{GR1v}\\
& &
\frac{h}{f}\frac{f'}{r}+
\frac{h-1}{r^2}=\frac{8\pi G P}{c^4}\,,
\label{GR2v}
\ea
which are exactly the same as Eqs.~(\ref{GR1}) and 
(\ref{GR2}) in GR, respectively. 
This shows that the intrinsic vector-mode couplings 
do not give rise to any modifications to the metric 
components $f$ and $h$. 
The TOV equation also holds in the same form as
Eq.~(\ref{GR4}). Thus, for a given EOS, the solutions 
to $f,h,P,\rho$ are the same as those in GR with 
$A_0(r)={\rm constant}$.
Requiring the smooth matching of the metric and vector field 
at the surface, Eq.~(\ref{A0const}) remains the solution outside 
the star with the exterior metric given by 
the Schwarzschild solution \eqref{Sch}.

The above property is in stark contrast with that in 
cubic and quartic couplings where the differential equation 
corresponding to Eq.~(\ref{A0dev}) contains the 
$A_0$-dependent terms which are not multiplied 
by the powers of $A_0'$. 
As we discussed in Secs.~\ref{sec4} and \ref{sec5},  
the existence of such terms leads to the variation of 
$A_0(r)$ for $r>0$. 
We also note that the presence of mass contribution 
$m^2 X$ to $G_2$ gives rise to the terms 
$m^2A_1$ and  $m^2 A_0$ to Eqs.~(\ref{A1eqv}) and 
(\ref{A0dev}), respectively, so the general solution to 
$A_0(r)$ is different from Eq.~(\ref{A0const}). 

In summary, the intrinsic vector-mode couplings (\ref{invec})
only lead to the metric components in GR with the trivial temporal 
vector component (\ref{A0const}) as the unique solution 
for relativistic stars, indicating no-hair properties unlike the 
BH solutions studied in Refs.~\cite{HKMT,HKMT2}. 
This no-hair property of relativistic stars is intrinsically 
related to the regular boundary condition $A_0'(r)=0$ at the center of star together with the peculiar structure of the differential Eq.~(\ref{A0dev}).
The result in this section holds irrespective of the choice of the 
coupling functions and the detail of EOSs.

\section{Conclusions}
\label{sec7}

In this paper, we studied how the mass-radius relation of 
relativistic stars 
is modified in generalized Proca theories. 
In these theories there exists a $U(1)$-breaking vector field 
with derivative couplings, which leads to the propagation of 
fifth forces. On the weak gravitational background in Solar System, it is known that such fifth forces can be suppressed by derivative self-interactions under the operation of the Vainshtein mechanism \cite{DeFeliceVain,Nakamura:2017lsf}.
On the other hand, the deviation from GR can manifest itself 
in the strong gravitational regime like 
BHs \cite{HKMT,HKMT2}.
Indeed, there exist a bunch of hairy BH solutions in generalized 
Proca theories. Our interest in this paper was to show how the 
new ``hair'' induced by vector-field derivative couplings affects 
the configuration of relativistic stars.

In Sec.~\ref{sec4} we considered the cubic power-law derivative 
coupling (\ref{cuG3}) including the vector Galileon ($n=1$) 
as a specific case. In these models, the vector field has a 
nonvanishing longitudinal mode $A_1$ related to the 
temporal component $A_0$ according to Eq.~(\ref{A1be3}). 
Imposing the regularity of metrics, pressure, density, and vector field at the center of star ($r=0$), we derived the analytic solutions 
(\ref{G3ini1})-(\ref{G3ini4}) around $r=0$. 
As we see in Eq.~(\ref{G3ini4}), the negative coupling constant 
$\beta_3$ leads to a slower decrease of the matter pressure $P(r)$.
This slower decrease continues up to the star surface, so the 
resulting radius $R_*$ for $\beta_3<0$ tends to be 
larger than that in GR. We also showed that the amplitude of 
negative coupling $\beta_3$ is constrained as Eq.~(\ref{bebo3})
from the demand $M(r)>0$ around $r=0$. 
This limits the maximum radius reached by the cubic coupling,  
see Eq.~(\ref{Rstar}). These properties hold independently 
of the EOS of relativistic stars.
 
To compute the mass $M_{\ast}$ and the radius $R_{\ast}$
of relativistic stars precisely, we numerically solved 
Eqs.~(\ref{be1})-(\ref{be5}) for the cubic Galileon coupling 
$G_3=\beta_3 X$ by employing the polytropic EOS (\ref{eos}) with $\Gamma=2.34$.
We confirmed that the negative coupling $\beta_3$ 
gives rise to $R_{\ast}$ larger than in the case $\beta_3=0$.
Although the mass function $M(r)$ is suppressed by negative  
$\beta_3$ around $r=0$, the increase of $R_{\ast}$ overwhelms 
this decrease to realize the mass $M_{\ast}$ greater than 
that in GR.  As we observe in Fig.~\ref{fig2}, the maximum 
mass $M_{\rm max}$ 
increases for a larger temporal vector component $a_0$ at $r=0$ 
and for an increasing amplitude of negative coupling $\beta_3$.
For $\beta_3>0$, both $M_\ast$ and $R_\ast$ are smaller 
than those in GR. Moreover, the models with large positive 
values of $\beta_3$ and $a_0$ are prone to 
instabilities associated with a negative gravitational binding 
energy $\Delta$ in the low-density regime.

In Sec.~\ref{sec5} we studied the effect of quartic power-law couplings (\ref{G4}) on the configuration of relativistic stars 
by considering the branch $A_1=0$. 
Again, the negative coupling $\beta_4$ leads to 
the larger mass $M_{\ast}$ and the larger radius $R_{\ast}$
relative to those in GR.
The difference from cubic derivative interactions is that the amplitude 
of negative $\beta_4$ is not constrained from the condition 
$M(r)>0$. For $\beta_4>0$ we found that both 
$M_{\ast}$ and $R_{\ast}$
are smaller than those in GR for most of the parameter space. 
The solutions are also subject to instabilities 
in the low-density regime with increasing values of 
$\beta_4$ and $a_0$. 
This is not the case for negative $\beta_4$ where the 
necessary condition for the dynamical stability is satisfied.

In Sec.~\ref{sec6} we showed that the intrinsic vector-mode couplings 
(\ref{invec}) give rise to solutions same as those in GR with the constant value of $A_0$. This is attributed to the peculiar structure of the 
differential equation (\ref{A0dev}) as well as the regular boundary 
condition $A_0'=0$ at $r=0$. 
Thus, the intrinsic vector modes do not modify the radius 
and mass of relativistic stars in GR.

There are several issues we did not address in this paper. 
We adopted the polytropic EOS (\ref{eos}) 
with $\Gamma=2.34$ to compute the mass and radius of relativistic stars, but for the comparison of them
with the observational data of NSs, 
we need to extend the analysis 
to more realistic EOSs 
by taking into account nuclear interactions and the 
composition of each layer of NSs.
It is also possible to include the rotation of NSs 
in our analysis along the line of Ref.~\cite{Hartle1}
and investigate the existence of EOS-independent 
relations \cite{Doneva2017} useful to test generalized Proca theories 
with NSs further.
Although we have confirmed that most of the solutions 
obtained in this paper are gravitationally bound,
the analysis of dynamical stabilities against odd- and
even-parity perturbations 
may provide further constraints on 
couplings in generalized Proca theories.
With this perturbative analysis on the spherically symmetric 
background, we should also be able to derive the local propagation 
speed $c_g$ of gravitational waves around NSs.
If the vector-field derivative couplings studied in this paper 
are also responsible for today's cosmic acceleration, 
the recent GW170817 bound of $c_g$ \cite{LIGO2} 
on the cosmological background
will provide tight constraints on quartic derivative couplings. 
These interesting issues will be left for future works.


\appendix

\section{
Coefficients in the gravitational equations of motion}
\label{app}

The coefficients $c_{1,2,\cdots,19}$ in 
Eqs.~(\ref{be1}) and (\ref{be2}) are given by 
\bea
c_{1} &=& -A_{1} X G_{3,X},
\\
c_{2} &=& -2 G_{4} + 4 (X_{0} + 2 X_{1}) G_{4,X} + 8 X_{1} X G_{4,XX},
\\
c_{3} &=& -A_{1} (3 h X_{0} + 5 h X_{1} - X) G_{5,X} - 2 h A_{1} X_{1} X G_{5,XX},
\\
c_{4} &=& G_{2} - 2 X_{0} G_{2,X} - \frac{h}{f} (A_{0} A_{1} A_{0}' + 2 f X A_{1}') G_{3,X} 
-\frac{h A'^{2}_{0}(1+2 G_{2,F})}{2 f}\,,
\\
c_{5} &=& -4 h A_{1} X_{0} G_{3,X} - 4 h^{2} A_{1} A'_{1} G_{4,X} 
+ \frac{8 h}{f} \left( A_{0} X_{1} A'_{0} - f h A_{1} X A'_{1} \right) G_{4,XX}
+\frac{2 h^{2}}{f} A_{1} A'^{2}_{0} (g_{5} + 2 X_{0} g_{5,X}),
\\
c_{6} &=& 2 (1 - h) G_{4} + 4 (h X - X_{0}) G_{4,X} +8 h X_{0} X_{1} G_{4,XX} 
- \frac{h}{f} \left[ (h - 1) A_{0} A_{1} A'_{0} + 2 f (3 h X_{1} + h X_{0} - X) 
A'_{1} \right] G_{5,X}
\\&&-\frac{2 h^{2} X_{1}}{f} (A_{0} A_{1} A'_{0} + 2 f X A'_{1}) G_{5,XX} 
+ \frac{h A'^{2}_{0}}{f} \left[ (h - 1) G_{6} + 2 (h X - X_{0}) G_{6,X} + 4 h X_{0} X_{1} G_{6,XX} \right],
\\
c_{7} &=& -G_{2} + 2 X_{1} G_{2,X} - \frac{h}{f} A_{0} A_{1} A'_{0} G_{3,X}
+\frac{h A'^{2}_{0}(1+2 G_{2,F})}{2 f}\,,
\\
c_{8} &=& 4 h A_{1} X_{1} G_{3,X} + \frac{4 h}{f} A_{0} A'_{0} (G_{4,X}+2 X_{1} G_{4,XX})
-\frac{2 h^{2}}{f} A_{1} A'^{2}_{0} (3 g_{5} + 2 X_{1} g_{5,X})\,, 
\\
c_{9} &=& 2 (h - 1) G_{4} - 4 (2 h - 1) X_{1} G_{4,X} - 8 h X_{1}^{2} G_{4,XX} 
- \frac{h}{f} A_{0} A_{1} A'_{0}\left[ (3 h -1) G_{5,X} + 2 h X_{1} G_{5,XX} \right] 
\\&&-\frac{h}{f} A'^{2}_{0} \left[ (3 h - 1) G_{6} + 2 (6 h - 1) X_{1} G_{6,X} + 4 h X_{1}^{2} G_{6,XX} \right]\,.
\eea
\section*{Acknowledgements}

We thank Lavinia Heisenberg for useful discussions.
RK is supported by the Grant-in-Aid for Young Scientists B 
of the JSPS No.\,17K14297. 
MM is supported by FCT-Portugal through 
Grant No.\ SFRH/BPD/88299/2012. 
ST is supported by the Grant-in-Aid for Scientific Research Fund of the JSPS No.~16K05359 and 
MEXT KAKENHI Grant-in-Aid for 
Scientific Research on Innovative Areas ``Cosmic Acceleration'' (No.\,15H05890).


\end{document}